\documentclass[aps,11pt]{revtex4}
\pdfoutput=1
\usepackage{bm}
\usepackage{times}
\usepackage{graphicx}
\usepackage{cancel}
\usepackage{amsmath}
\usepackage{amssymb}
\usepackage{textcomp}
\usepackage{color}

\def\beq{\begin{equation}}
\def\eeq{\end{equation}}
\def\be{\begin{equation}}
\def\ee{\end{equation}}
\def\bea{\begin{eqnarray}}
\def\eea{\end{eqnarray}}
\renewcommand{\eqref}[1]{Eq.~(\ref{#1})}

\def\ev{\,{\rm eV}}

\newcommand{\cf}{cf.\/}
\newcommand{\eg}{e.\,g.\/}
\newcommand{\ie}{i.\,e.\/}
\def \mm{\mathcal{M}}
\def\lsim{\mathrel{\rlap{\lower4pt\hbox{\hskip1pt$\sim$}}\raise1pt\hbox{$<$}}}
\def\gsim{\mathrel{\rlap{\lower4pt\hbox{\hskip1pt$\sim$}}\raise1pt\hbox{$>$}}}
\def\BR{{\rm BR}}

\newcommand{\mFal}{m_{F_{\alpha}}}
\newcommand{\mSpm}{m_{S^{\pm}}}
\newcommand{\mSR}{m_{S^{0}_R}}
\newcommand{\mSI}{m_{S^{0}_I}}
\newcommand{\mSRI}{m_{S^{0}_{R/I}}}
\newcommand{\qqbar}{q\bar{q}}

\newcommand\plot[2]{\includegraphics[width=#1\linewidth, angle=0]{plots_Oct25/#2}}
\newcommand\Fdiag[2]{\includegraphics[#1]{FeynDiags_Oct01/#2}}

\begin{document}
%
\begin{flushright}
MADPH-10-1566 
\end{flushright}
%

\title{\Large Lepton Number Violation from Colored States at the LHC}
\author{Pavel Fileviez P\'erez}
\author{Tao Han}
\author{Sogee Spinner}
\author{Maike K. Trenkel\footnote{email addresses: 
fileviez@physics.wisc.edu,~ than@hep.wisc.edu,~ sspinner@wisc.edu,~ trenkel@hep.wisc.edu.  } }
\affiliation{University of Wisconsin-Madison, Department of Physics \\
1150 University Avenue, Madison, WI 53706, USA}

\date{\today}

\begin{abstract}
The possibility to search for lepton number violating signals at the Large Hadron Collider (LHC) in the 
colored seesaw scenario is investigated. In this context the fields that generate neutrino masses at the 
one-loop level are scalar and Majorana fermionic color-octets of $SU(3)_{C}$. Due to the QCD strong 
interaction these states may be produced at the LHC with a favorable rate. We study the production 
mechanisms and decays relevant to search for lepton number violation signals in the channels with 
same-sign dileptons. In the simplest case when the two fermionic color-octets are degenerate in mass, one 
could use their decays to distinguish between the neutrino spectra. We find that for fermionic octets with 
mass up to about 1 TeV the number of same-sign dilepton events is larger than the standard model 
background indicating a promising signal for new physics.  
\end{abstract}

\maketitle

\section{Introduction}
The LHC Era brings with it the hope of the discovery of the New Standard Model of particle physics. 
Searching for lepton and/or baryon number violation at the LHC might be an integral part of establishing 
this new theory. As it is well known, in the Standard Model (SM) the lepton and baryon numbers are 
accidental global symmetries at the classical level. Typically, beyond the SM physics scenarios introduce 
new interactions which can violate these symmetries. A concrete example are Majorana neutrinos, 
which require extra states and interactions to generate neutrino masses. These can give rise to new and 
interesting phenomenology relevant for the LHC.

Recently, a simple mechanism for the generation of Majorana neutrino masses at one-loop level was proposed 
where the new fields, inside the loop, live in the adjoint representation of $SU(3)_C$~\cite{Perez-Wise}. 
In this context two types of new fields are required: a fermionic octet, $\rho \sim (8,1,0)$, and a scalar octet 
$S \sim (8,2,1/2)$. It is important to stress that this is the simplest mechanism at one loop allowed by 
cosmology which does not require an extra symmetry. We refer to this model as the ``colored seesaw mechanism". 
This is a novel and interesting possibility which one can use to test the origin of neutrino masses and look for lepton 
number violating signals at the LHC since the seesaw fields can be easily produced via the strong interaction.

In this article, we investigate these lepton number violating signals at the LHC in the context of the colored seesaw 
mechanism. We study the production mechanisms and decays of the colored states, focusing on same-sign dileptons in the final state as an indicator of lepton number violation. In the simplest case when the two fermionic color-octets are 
degenerate in mass, their decays allow us to distinguish between the neutrino spectra. We find that for 
fermionic octets with mass up to a few TeV the number of same-sign dilepton events is larger than the SM background.  While our study is only on the level of cross section times branching ratio, it is still a promising result for the testability of this mechanism at the LHC.

The paper is organized as follows: In Section II we set the stage for this paper by summarizing the potential 
for observing lepton number violation in the simplest extensions of the SM with Majorana neutrino masses. 
In Section III we describe in detail the colored seesaw mechanism and the constraints coming from neutrino 
physics and lepton flavor violation. Section IV is devoted to the discussion of the testability of the model 
at the LHC: we discuss the possible two- and three-body decays, and the production of the fermionic octets.
We finish by summarizing our findings in Section V. 

\section{Lepton Number Violation at the LHC}
It is well known that neutrinos are either Dirac or Majorana fermions. Majorana neutrinos allow 
for a great variety of scenarios in which neutrino masses originate from integrating out heavy fields, the generic seesaw 
mechanism. A tantalizing possibility is that the new heavy fields could be produced at the LHC and, due to their Majorana
nature, give rise to lepton number violating signals.   

At tree level, neutrino masses can be generated through the well-known Type I, Type II or Type III seesaw mechanisms:

\begin{itemize}

\item \textit{Type I seesaw mechanism}~\cite{TypeI}: The SM is extended by at least two SM singlets, $\nu^C \sim (1,1,0)$.  
Once those singlets are integrated out the neutrino mass matrix reads as ${\cal M}^I_\nu = Y_\nu \  M_R^{-1} \ Y_\nu^T v^2$, 
where $Y_\nu$ is the Yukawa coupling between the SM leptonic doublet and the right-handed neutrinos, $v=246$ GeV is the 
vacuum expectation value of the SM Higgs boson, and $M_R$ is the Majorana mass matrix for the right handed neutrinos. It is difficult 
to look for lepton number violation in this case due to the singlet nature of the right-handed neutrinos, which does not allow them 
to be easily produced. See Ref.~\cite{TypeI-LHC} for details.

\item \textit{Type II seesaw mechanism}~\cite{TypeII}:  An $SU(2)$ scalar triplet is introduced, $\Delta \sim (1,3,1)$, and the 
neutrino masss matrix is given by ${\cal M}^{II}_\nu =  h_\nu \ v_{\Delta}$. Here, $h_\nu$ is the Yukawa coupling between 
the leptons and the triplet, and $v_{\Delta}$ is the vacuum expectation value of the neutral component of the triplet. In this 
scenario one could find spectacular signals at the LHC from the decays of singly and doubly charged components of the Higgs 
triplet into leptons, i.e. $H^{\pm} \to e^{\pm} \nu $ and $H^{\pm \pm} \to e^{\pm}_i e^{\pm}_j$~\cite{TypeII-LHC}. 

\item \textit{Type III seesaw mechanism}~\cite{TypeIII}: It is also possible to generate neutrino masses at tree level by introducing
at least two extra fermions in the adjoint representation of $SU(2)$, $\rho \sim (1,3,0)$. The mass matrix for neutrinos is similar to the 
Type I case, where one replaces $M_R$ by $M_\rho$,  the Majorana mass matrix for the fermionic triplets. For the testability of the 
Type III seesaw at the LHC see~\cite{TypeIII-LHC}.

\end{itemize}
These are the simplest mechanisms for generating neutrino masses at tree level since they involve only the addition of one new representation 
to the minimal Standard Model. It is important to mention that all these scenarios can be realized in the context of grand unified 
theories (GUT's) based on $SU(5)$ and $SO(10)$ gauge symmetries. The realization of the Type III seesaw mechanism in the context 
of GUT's is special since it always leads to a hybrid scenario: Type I plus Type III seesaw~\cite{TypeIII,Bajc-Senjanovic,Adjoint}. 
See Ref.~\cite{PFP-review} for a review of the different seesaw mechanisms. 

The simplest mechanisms for the generation of neutrino masses at one-loop level are

\begin{figure}[bt]
\Fdiag{}{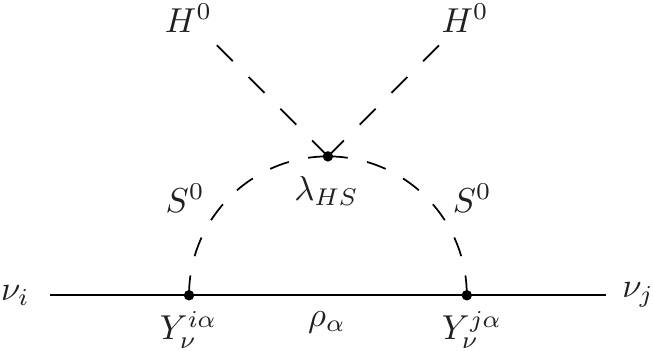}
\caption{Feynman diagram for neutrino mass generation in the colored seesaw mechanism~\cite{Perez-Wise}
introducing a scalar colored octet $S \sim (8,2,1/2)$ and two fermionic octets $\rho_{\alpha} \sim (8,1,0)$.
$\lambda_{HS}$ is the quartic scalar-Higgs coupling and $Y_{\nu}$ the neutrino Yukawa couplings.
\label{fig_seesaw}}
\end{figure}

\begin{itemize}

\item \textit{Zee-Mechanism}~\cite{Zee}: Neutrino masses can be generated at one-loop level with the addition of two extra Higgs bosons: 
a Higgs singlet $\delta \sim (1,1,1)$ and a Higgs doublet $H_2 \sim (1,2,1/2)$. In this scenario it is difficult to look for lepton number 
violation because one has only singly charged Higgs bosons which do not allow for the like-sign dilepton signals.

\item \textit{Colored Seesaw}~\cite{Perez-Wise}:  In this mechanism the ``seesaw" fields live in the adjoint representation 
of $SU(3)$. Adding one type of fermionic representation, $\rho \sim (8,1,0)$, and one colored scalar, $S \sim (8,2,1/2)$, it is possible 
to generate neutrino masses at the one-loop level, as shown in Figure~\ref{fig_seesaw}. If the fermionic fields are light, with mass 
below or around the TeV scale, one can hope to search for lepton number violation through the production of colored states 
which can be easily produced via the strong interactions at the LHC. This is a novel and new possibility which we will explore in this paper. 

\end{itemize}
It is important to mention that in the context of supersymmetric theories with R-parity violation one has many interesting predictions 
for lepton number violating decays which can be found at the LHC. For a review see Ref.~\cite{R-review}. 


\section{Colored Seesaw Mechanism}
The Colored Seesaw Mechanism~\cite{Perez-Wise} introduces two new types of fields: a scalar color octet $S \sim (8,2,1/2)$ 
and a fermionic octet $\rho \sim (8,1,0)$. Neutrino masses are generated at the one-loop level as shown in Fig.~\ref{fig_seesaw}. 
It is important to emphasize that this model is the simplest one of this type at the one-loop level which does not require an extra symmetry. 
The scalar octet has the same quantum numbers, aside from color, as the SM Higgs boson therefore leading to a quartic interactions between the two fields and couplings to the Standard Model quark 
fields~\cite{Manohar}; the $\rho$ has the same quantum numbers as the gluino in supersymmetric models. 

The relevant interactions (here we write just one possible quartic interaction for simplicity) needed to generate neutrino masses are given by,
\begin{equation}
-{\cal L}_\nu = Y_\nu \ l^T \ C \ i \sigma_2 \ S \  \rho \ + \frac{1}{2} \ M_{\rho} \ {\rm Tr} \left(\rho^T \ C \ \rho\right) \ 
	+ \ \lambda_{HS} \  {\rm Tr} \left( S^\dagger \ H \right)^2 \ + \ {\rm h.c.} ~,
\label{V2}
\end{equation}
where $Y_{\nu}$ are the lepton doublet Yukawa couplings to the seesaw fields, $\lambda_{HS}$ is the quartic scalar-Higgs coupling, and 
$\sigma_2$ is a Pauli matrix and $C$ the charge conjugation operator.  Integrating out the seesaw fields leads to the usual dimension five operator for neutrino masses. All fermions have left chirality, $l^T=(\nu, \ e)$, $S^T=(S^+, S^0)$, and the trace is over color indices.
In order to generate neutrino masses in agreement with experiments at least two copies of  $S$ or $\rho$ are needed.
Notice that this scenario is consistent cosmologically since the scalar octet can decay into SM quarks ~\cite{Perez-Wise}. 
Finally, we would like to point out that this mechanism can be realized in the context of the Adjoint $SU(5)$~\cite{Adjoint} 
grand unified theory since $\rho \subset 24$ and $S \subset 45_H$.
 
In the following we focus on the simplest model with two copies of the new fermions and one scalar colored octet. 
Working in the mass eigenstate basis for the two new fermions $\rho_{\alpha}$, $\alpha = 1,2$,  the $3\times3$ neutrino 
mass matrix reads as,
\begin{equation}
\label{mnuformula}
{\cal M}_\nu^{ij}=Y_{\nu}^{i \alpha} \ Y_{\nu}^{j \alpha} \  \frac{\lambda_{HS}}{16 \pi^2} \  v^2 \
 I\left(m_{\rho_\alpha}, m_{S}\right), 
\end{equation}
where the loop integration factor,  $I \left(m_{\rho_\alpha}, m_{S}\right)$, is given by
\begin{equation}
\label{loop.integration}
I \left( m_{\rho_\alpha}, m_{S}\right) = m_{\rho_\alpha}  
	\frac{\left( m_{S}^2-m_{\rho_\alpha}^2+m_{\rho_\alpha}^2{\rm ln}(m_{\rho_\alpha}^2/m_{S}^2) \right)}
		{\left(m_{S}^2- m_{\rho_\alpha}^2\right)^2 }.
\end{equation}
With this minimal number of new fields there are two massive and one massless neutrino.  
If we stick to the maximal suppresion coming from the loop factor, the scalar mass should be larger than the fermionic mass.
This scenario provides the best chance to observe lepton number violation since the Yukawa coupling could be large.
In the limit  $m_{S} \gg m_{\rho}$ the loop factor simplifies and the neutrino mass matrix becomes,
\begin{equation}
\label{mnuformula1}
{\cal M}_\nu^{ij}=Y_{\nu}^{i \alpha} \ Y_{\nu}^{j \alpha} \  \frac{\lambda_{HS}}{4 \pi^2} \  v^2 \ \frac{m_{\rho_\alpha}}{m_{S}^2}.
\end{equation}
For illustration purposes let us consider $m_{\rho} = 200$ GeV and $m_{S}=2$ TeV. With $v=246$ GeV, we find that in order 
to get the neutrino mass ``scale", $\sim 1$ eV, the combination of the couplings has to be $Y_{\nu}^2 \lambda_{HS} \sim 10^{-8}$. 
Notice that the simultaneous presence of the Yukawa term proportional to $Y_\nu$ and the quartic interaction proportional 
to $\lambda_{HS}$ in Eq.~(\ref{V2}) violate lepton number. Lepton flavor conservation is also violated, 
even when $\lambda_{HS}$ is very small, by the Yukawa couplings $Y_{\nu}$ through processes such as
$\mu \rightarrow e +\gamma$~\cite{Tulin, Liao}. We will comment on this in Section \ref{subsec_brmu2egamma}.

\subsection{Properties of the Scalar Octet}
The extra Yukawa 
interactions due to the presence of the scalar octet $S$ are given by
\begin{equation}
{\cal L}_Y = \bar{d}_R \ \Gamma_D \ S^\dagger \ Q_L \ + \ \bar{u}_R \ \Gamma_U \ Q_L^x \ S^y \ \epsilon_{xy} \ + \ \text{h.c.},
\label{int1}
\end{equation}
where $x,y$ are $SU(2)$ indices and
\begin{equation}
S =
\left(
\begin{array} {c}
 S^+  \\
 S^0
\end{array} \right)
=
\left(
\begin{array} {c}
 S^+  \\
 \frac{1}{\sqrt{2}} (S_R^0 + i S_I^0)
\end{array} \right)
=S^A T^A,
\end{equation}
with $A=1,\ldots,8$ and $T^A$ being the $SU(3)$ generators.
In the physical basis the Lagrangian reads,
\begin{align}
\begin{split}
\mathcal L_Y=
&\bar d\left[P_L\left(D_R^\dagger\Gamma_DU_L\right)-P_R\left(D_L^\dagger\Gamma_U^\dagger U_R\right)\right]S^-u+\bar u\left[P_R\left(U_L^\dagger\Gamma_D^\dagger D_R\right)-P_L\left(U_R^\dagger\Gamma_U D_L\right)\right]S^+d
\\
&+\frac{S_R^0}{\sqrt2}\bar d\left[P_L\left(D_R^\dagger\Gamma_D D_L\right)+P_R\left(D_L^\dagger\Gamma_D^\dagger D_R\right)\right]d+\frac{S_R^0}{\sqrt2}\bar u\left[P_L\left(U_R^\dagger\Gamma_U U_L\right)+P_R\left(U_L^\dagger\Gamma_U^\dagger U_R\right)\right]u
\\
&-i\frac{S_I^0}{\sqrt2}\bar d\left[P_L\left(D_R^\dagger\Gamma_DD_L\right) - P_R\left(D_L^\dagger\Gamma_D^\dagger D_R\right)\right]d+i\frac{S_I^0}{\sqrt2}\bar u\left[P_L\left(U_R^\dagger\Gamma_U U_L\right)-P_R\left(U_L^\dagger\Gamma_U^\dagger U_R\right)\right]u.
\end{split}
\end{align}
Here $u$ and $d$ are respectively the SM up- and down-type quark fields and 
  $U_L$, $U_R$, $D_L$ and $D_R$ are the matrices that diagonalize the quark mass matrices.  $S^{\pm}$ 
denotes the charged octet scalar and $S^0_{R,I}$ are respectively the $CP$-even and $CP$-odd neutral scalars. 
Assuming minimal flavor violation~\cite{Manohar}, which we do for the rest of this work, means that
\begin{equation}
\Gamma_U = \eta_U Y_U \;\text{and} \;\Gamma_D = \eta_D Y_D.
\label{etadef}
\end{equation}
In this case the physical interactions are
\begin{align}\label{lagrangian}
\begin{split}
\mathcal L_Y^\text{MFV}=
&\frac{\sqrt2}v\bar d\left(
	P_L\eta_D\ m_DV_\text{CKM}^\dagger 
	-P_R\eta_U\ V_\text{CKM}^\dagger m_U
\right)S^-u
\\
&+\frac{\sqrt2}v\bar u\left(
	P_R\eta_D\ V_\text{CKM}m_D
	-P_L\eta_U\ m_UV_\text{CKM}
\right)S^+d
\\
&	+\eta_D\frac{m_D}v\ S_R^0\bar dd
	+\eta_U\frac{m_U}v\ S_R^0\bar uu
	+i\eta_D\frac{m_D}v\ S_I^0\bar d\gamma_5 d
	-i\eta_U\frac{m_U}v\ S_I^0\bar u\gamma_5 u,
\end{split}
\end{align}
in terms of the quark masses  $m_U,m_D$ and  the Cabibbo-Kobayashi-Maskawa matrix  $V_\text{CKM}$. 
$\eta_U$ and $\eta_D$ are parameters that describe the strength of the scalar couplings to matter. 
For a list of the Feynman rules see Fig.~\ref{fig_Feynman_S} in Appendix~\ref{sect_FeynmanRules}.

The masses of the charged state and the neutral members of the scalar octet depend on the details of the potential describing
their self-interactions. The vacuum expectation value of the SM Higgs boson causes a mass splitting between the 
octet scalars~\cite{Manohar}. If the splitting is sufficiently large, decays such as $S^{\pm} \to S^0 W^{\pm}$ are allowed, 
where $S^0$ denotes one of the neutral scalars. However, the mass squared splitting is on the order of the electroweak scale 
squared and will only lead to significant mass splittings for light scalars, which are disfavored due to their contribution 
to $b \to s \gamma$. Therefore, it is likely that the charged scalar will decay predominantly via $S^{+} \to t\bar{b}$ due 
to the couplings in Eq.~(\ref{lagrangian}) and we adopt this assumption for the rest of this study. See Ref.~\cite{Octets} 
for the study of the production mechanisms at the LHC and the properties of their decays.
\subsection{Properties of the Fermionic Octets}
In order to derive the Feynman rules for the fermionic octet we rewrite Eq.~(\ref{V2}) in a more traditional form:
\begin{equation}
	-\mathcal{L}_\nu \supset
		Y_\nu^{i \alpha} \ \bar l_{L_i} \ i \sigma_2
		\text{Tr}\left(S^\dagger \ \rho_{R_\alpha}\right)
		\ + \frac{1}{2} \ M_\rho^{\alpha} \ \text{Tr} \left( \rho^T_{R_\alpha} \ C \ \rho_{R_\alpha}\right)
		\ + \ \text{h.c.},
\end{equation}
where the fermionic mass matrix $M_\rho$ can be assumed to be diagonal without loss of generality. 
Now, defining a Majorana fermion as 
\begin{equation}
	F \equiv
	\begin{pmatrix}
		(\rho_R)^C
		\\
		\rho_R
	\end{pmatrix},
\end{equation}
the four-component Lagrangian with the mass eigenstates of $S$ has the form:
\begin{eqnarray}
	-\mathcal{L}_\nu &\supset &
		\frac{1}{\sqrt 2} \ \left(V^T_{\rm PMNS} \ Y_\nu \right)_{i\alpha} \bar \nu_i \ P_R \ \text{Tr} \left(S^{0}_R \ F_\alpha \right)
		\ + \ \frac{1}{\sqrt 2} \ \left(Y_\nu^\dagger \ V^*_{\rm PMNS}\right)_{\alpha i} \text{Tr} \left(S^{0}_R \ \bar F_\alpha \right) \ P_L \ \nu_i
		\nonumber\\
		& - & \frac{i}{\sqrt 2} \ \left(V^T_{\rm PMNS} \ Y_\nu\right)_{i\alpha} \bar \nu_i \ P_R \ \text{Tr} \left(S^0_I\ F_\alpha \right)
		\ + \ \frac{i}{\sqrt 2} \ \left(Y_\nu^\dagger \ V^*_{\rm PMNS}\right)_{\alpha i} \text{Tr} \left(S^{0}_I \ \bar F_\alpha \right) \ P_L \ \nu_j
		\nonumber\\
		& - & Y_\nu^{i \alpha} \ \bar l^-_i \ P_R \ \text{Tr} \left(S^{-} \ F_\alpha\right)
		\ - \ Y_\nu^{\alpha i *} \ \text{Tr}\left(S^{+} \ \bar F_\alpha\right) \ P_L \ l^-_i
		\ +  \frac{1}{2} M_\rho^\alpha \ \text{Tr} \ \bar F_\alpha \ F_\alpha.
\end{eqnarray}
The corresponding Feynman rules are given in Fig.~\ref{fig_Feynman_S} in Appendix~\ref{sect_FeynmanRules}.
 
We note that collider experiments allow to set a conservative lower bound on the mass of the fermionic octets of about 200 GeV, 
as for the gluino in supersymmetric theories.

\subsection{Constraints from Neutrino Physics}
\label{subsec_neutrinoconstraints}
%
Since the properties of the fermionic octets depend on the neutrino Yukawa couplings and thus on 
the neutrino mass hierarchy, it is helpful to express 
the relevant Yukawa couplings as a function of the leptonic mixing angles and neutrino masses. 
Starting from Eq.~(\ref{mnuformula}), we write the neutrino mass matrix $\mm_\nu$ as
\begin{equation}
\mm_\nu = Y_\nu \  (M^{\rm Eff})^{-1} \ Y_\nu^T \ v^2,
\end{equation} 
where $M^{\rm Eff}$ is a $2\times2$ diagonal matrix,
\begin{equation}
M^{\rm Eff}  = 
 \frac{16 \pi^2}{\lambda_{HS}} \ {\rm diag}(I\left(m_{\rho_1}, m_S \right), I\left(m_{\rho_2}, m_S \right))^{-1}.
\label{eq_Meff}
\end{equation} 
and $I\left(m_{\rho_\alpha}, m_S \right)$ is the loop integration function expressed in Eq.~(\ref{loop.integration}). 
The three light neutrino masses can be expressed in the following way,
\begin{equation}
m_{\nu} = V_{\rm PMNS}^\dagger \, \mm_\nu \ V_{\rm PMNS}^*,
\end{equation}
where $m_{\nu}={\rm diag} (m_1, m_2, m_3)$ and $V_{\rm PMNS}$ can be taken as  the
leptonic mixing matrix for the three generation of light neutrinos without the loss of generality.
Working in the basis where the charged lepton mass matrix is diagonal, one finds for the neutrino Yukawa couplings $Y_{\nu}$,
\begin{equation}
Y_{\nu}= \frac{1}{v} \ V_{\rm PMNS} \ m_{\nu}^{1/2} \ \Omega \ (M^{\rm Eff})^{1/2}.
\label{Dirac}
\end{equation}
Here we are using the 
Casas-Ibarra parametrization, where $\Omega$ is a complex matrix  
satisfying the orthogonality condition $\Omega^T \Omega = 1$~\cite{Casas:2001sr}.
 
In order to understand the constraints coming from neutrino physics
let us discuss the relation between the neutrino masses and mixing.
The leptonic mixing matrix is given by
\begin{equation}
V_{\rm PMNS}= \left(
\begin{array}{lll}
 c_{12} c_{13} & c_{13} s_{12} & e^{-\text{i$\delta $}} s_{13}
   \\
 -c_{12} s_{13} s_{23} e^{\text{i$\delta $}}-c_{23} s_{12} &
   c_{12} c_{23}-e^{\text{i$\delta $}} s_{12} s_{13} s_{23} &
   c_{13} s_{23} \\
 s_{12} s_{23}-e^{\text{i$\delta $}} c_{12} c_{23} s_{13} &
   -c_{23} s_{12} s_{13} e^{\text{i$\delta $}}-c_{12} s_{23} &
   c_{13} c_{23}
\end{array}
\right)\times \text{diag} (1, e^{i \Phi/2}, 1)
\end{equation}
where $s_{ij}=\sin{\theta_{ij}}$, $c_{ij}=\cos{\theta_{ij}}$, $0 \le
\theta_{ij} \le \pi/2$ and $0 \le \delta, \Phi \le 2\pi$. The phase
$\delta$ is the Dirac CP phase, and $\Phi$ is the Majorana
phase. The experimental constraints on the neutrino masses and
mixing parameters, at $2\sigma$ level~\cite{Schwetz}, are
\begin{eqnarray}
\begin{split}
7.25 \times 10^{-5} \ev^2 \  < &~ \Delta m_{21}^2 ~ < \  8.11 \times 10^{-5} \ev^2, \\
2.18 \times 10^{-3} \ev^2 \  < &~ |\Delta m_{31}^2| ~< \  2.64 \times 10^{-3} \ev^2, \\
                   0.27 \  < &~ \sin^2{\theta_{12}}  < \  0.35, \\
                   0.39 \  < &~ \sin^2{\theta_{23}} <\  0.63, \\
                          &~ \sin^2{\theta_{13}}  <\  0.040,
\label{eq_lept_mixingparams}
\end{split}
\end{eqnarray}
and $\sum_{i} m_{i} < \ 1.2 \ \ev$ \cite{Concha}. 
Following the conventions, we denote the case $\Delta m_{31}^2 > 0$ as the normal neutrino mass hierarchy
and $\Delta m_{31}^2 < 0$ the inverted hierarchy, \ie we have
\begin{align}
\begin{split}
m_1 = 0, \quad m_2 &= \sqrt{\Delta m_{21}^2}, \quad m_3 = \sqrt{\left|\Delta m_{31}^2 \right|}
\qquad~~(\text{NH}),
\\ 
m_1 = \sqrt{\left|\Delta m_{31}^2 \right|}, \quad m_2 &= \sqrt{\left|\Delta m_{31}^2 \right|+ \Delta m_{21}^2 }, \quad m_3=0
\qquad (\text{IH}).
\end{split}
\end{align}
The $\Omega$ matrix takes the well-known form corresponding to the 
Type I seesaw case with two right-handed neutrinos \cite{ir}. Expressed in terms 
of an angle $\omega$, it reads
\begin{equation}\label{Omega}
\Omega^{\rm NH}=\left(
\begin{array}{cc}
  0  &  0   \\
  \sqrt{1- \omega^2} & - \omega \\
  \omega & \sqrt{1-\omega^2}
\end{array}
\right) ,\quad \Omega^{\rm IH}=\left(
\begin{array}{cc}
    \sqrt{1- \omega^2} & - \omega  \\
   \omega  & \sqrt{1- \omega^2}\\
0&0
\end{array}
\right),
\end{equation}
in the normal (NH) and inverted (IH) hierarchy, respectively. 
Here we focus on the range, $-1 \leq \omega \leq 1$, for simplicity.
%

\subsection{Constraints from Lepton Flavour Violation: $\mu \to e \gamma$}
\label{subsec_brmu2egamma}
\begin{figure}[tb]
\Fdiag{}{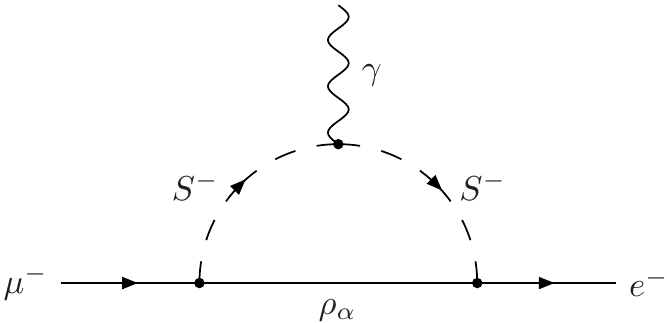}
\caption{Feynman diagram for the contribution to $\mu \to e \gamma$ in the colored seesaw model.}
\label{mu2egammaplot}
\end{figure}

\begin{figure}[bt]
\plot{.49}{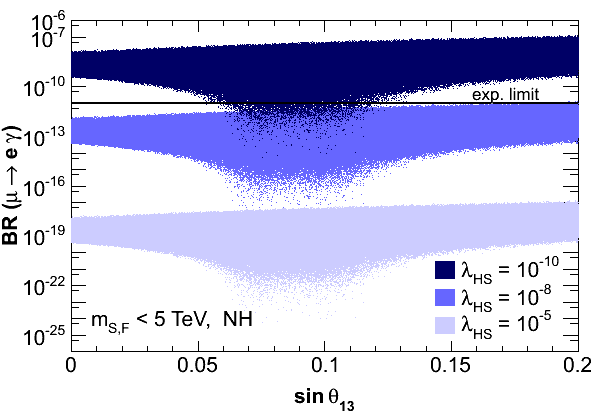}~
\plot{.49}{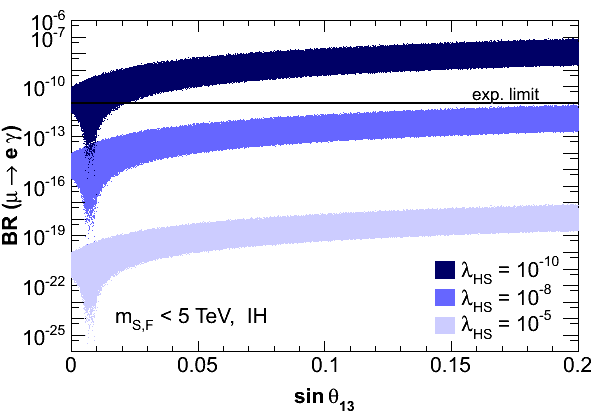}
\caption{ BR($\mu \to e \gamma$) as a function of $\sin\theta_{13}$ for
 1 TeV$ < m_S, m_F < $5 TeV with $m_{F_1} = m_{F_2}$.
In the left panel normal hierarchy (NH) and in the right panel inverted hierarchy (IH) is considered.
\label{fig_BRmu2egamma}
}
\end{figure}

Neutrino oscillations indicate the presence of lepton flavor violating operators 
which lead to various rare decays, the most stringently constrained being $\mu \to e \gamma$.  
The colored seesaw has a contribution which can potentially disagree with experiments 
(see Fig. \ref{mu2egammaplot}). 
The branching ratio for this process is given by
\begin{equation}
	\label{mu2egamma}
	\BR\left(\mu \to e \gamma \right) 
		= \frac{ \Gamma(\mu \to e \gamma)} {\Gamma(\mu \to e \nu \bar\nu)} 
		= \frac{3 \, \alpha_{EM}}{4 \, \pi \, G_F^2 \, m_S^4}
		\left|
			\displaystyle\sum_{\alpha=1}^{2} Y_\nu^{1\alpha} (Y_\nu^{\alpha 2})^* \ \mathcal{F}\left(x_\alpha\right)
		\right|^2,
\end{equation}
where $x_\alpha  =m_{\rho_\alpha}^2/ m_S^2$ and
\begin{align}
\mathcal{F}\left(x\right) & = \frac{1 - 6 \, x + 3 \, x^2 + 2 \, x^3 - 6 \, x^2 \, \ln x}{12 \left(x-1\right)^4}.
\end{align}
Here we have used $\Gamma(\mu \to e \nu \bar\nu) = G_F^2 m_{\mu}^2/(192 \pi^3)$.
 Notice that our result agrees with \cite{Liao} up to the normalization for the Yukawa couplings, but disagrees 
 with \cite{Tulin}. It is clear from the above that this branching ratio is very sensitive to the Yukawa couplings: $\BR\left(\mu \to e \gamma \right) \sim (Y_\nu)^4$.  Only for order one Yukawa couplings this process does conflict with the experimental bound of $\BR\left(\mu \to e \gamma \right) < 1.2 \times 10^{-11}$~\cite{mu2egamma}, which will be improved by two or three orders of magnitude in future experiments.  This corresponds to $\lambda_{HS} \sim 10^{-10}$ for the quartic coupling between the octet scalar and the SM Higgs.  While one can claim that this is inline with the spirit of the seesaw mechanism (order one Yukawa couplings) there is no argument for why the quartic coupling should be so small, although such a small value is protected by global 
 $U(1)_L$ lepton number.  Nevertheless, the interplay of the low energy constraints with the collider signatures can be exciting in this part of the parameter space.
In Fig.~\ref{fig_BRmu2egamma} we show the predictions for $\BR(\mu \to e \gamma)$ 
versus the values of  $\sin\theta_{13}$, for $1\,{\rm TeV} < m_{F_1} = m_{F_2}, m_S < 5\,{\rm TeV}$ 
in the case of NH and IH, respectively. One finds that the quartic coupling is restricted to
\begin{align}
	\lambda_{HS} \gsim 10^{-8}
\end{align}
in order to comply with the experimental bounds on the rare decay.
In the case of degenerate fermions, Eq.~(\ref{mu2egamma}) simplifies (the $\omega$ dependence drops out). 
Using Eq.~(\ref{Dirac}) to express the Yukawa couplings in terms of the neutrino parameters and the quartic coupling $\lambda_{HS}$ (see Appendix~\ref{Yukawa.Neutrino} for Yukawas explicitly given in terms of neutrino parameter) and expanding in $s_{13}$ shows that the rate in the NH goes as
\begin{equation}
	\BR\left(\mu \to e \gamma \right)  \propto \left| c_{12} \, c_{23} \, s_{12} \sqrt{\Delta m^2_{21}}
	+ s_{13} \, s_{23} \, e^{-i(\delta -\frac{\Phi}{2})} \sqrt{\Delta m^2_{31}} \right|^2  \frac{1}{\lambda_{HS}^2} . 
\end{equation}
This a sum of two terms which are both suppressed: the first by the solar mass parameter and the second by $s_{13}$.  This leads to a cancellation at around $s_{13} \sim 0.09$ and $\delta -\frac{\Phi}{2}=\pi$, corresponding to the dip in the left-hand panel of Fig.~\ref{fig_BRmu2egamma}.
In the case of IH the branching ratio goes as
\begin{equation}
	\BR\left(\mu \to e \gamma \right)  \propto \left| -s_{13} \, s_{23} \, e^{-i \, \delta} \sqrt{\Delta m^2_{31}}
	+ \frac{1}{2} \, c_{12} \, s_{12}  \, c_{23} \frac{\Delta m^2_{21}}{\sqrt{\Delta m^2_{31}}} \right|^2  \frac{1}{\lambda_{HS}^2} ,
\end{equation}
where a further expansion in the solar mass scale has been conducted.  Once more we have the sum of two suppressed terms with a zero at around $s_{13} \sim 0.007$ and $\delta=0$, again reflected in the right panel of Fig.~\ref{fig_BRmu2egamma}.
Since the neutrino Yukawa couplings define the decay length of the fermionic octets, as we will show in the next section, the above results thus define a lower bound on the decay length. 

%



\section{Colored Seesaw at the LHC}
In this article we focus on the pair production of the fermionic octets 
to understand the predictions for the lepton number violating 
channels with multileptons. The production mechanisms for scalar octets 
do not lead to lepton number violating signals and have been investigated in previous studies~\cite{Octets}.

\subsection{Decay Modes of the Fermionic Octets} 
Lepton number violation in the context of the colored seesaw model 
originates from the decays of the fermionic octets, $F_{\alpha}$.
Depending on the mass configuration of the octets, different decay channels dominate.  
If $\mFal > \mSpm$ (and $\mFal > \mSRI $), the dominant decay 
modes are two-body decays into a scalar octet state and a lepton,
\begin{align}
	F_{\alpha} \  \to 
	\left\{ \begin{array}{l}
		\ell^+_i \, S^-\\
		\ell^-_i \, S^+        
	\end{array}\right.  
\qquad\text{and}\qquad  
	F_{\alpha} \  \to 
	\left\{ \begin{array}{l}
		\bar{\nu}_i \, S^0_{R/I} \\
		\nu_i \, S^{0}_{R/I}       
	\end{array}\right. \, .
\label{eq_2body}
\end{align}
Here $\ell^{\pm}$ denotes a charged lepton and $i = 1,2,3$ is the generation index.
However, if $\mSpm > \mFal > m_t$, the fermionic octets predominantly 
decay via an off-shell (charged) scalar octet into a lepton and a heavy-flavor quark pair,
\begin{align}
	F_{\alpha} \  \to 
	\left\{ \begin{array}{l}
		\ell^+_i \, \bar{t} \, b \\
		\ell^-_i \, t \, \bar{b} 
	\end{array}\right.  \, .
\label{eq_3body_charged}
\end{align}
The three-body decay can also be mediated by a neutral scalar octet state, if $\mSRI > \mFal > 2m_t$,
\begin{align}
	F_{\alpha} \  \to 
	\left\{ \begin{array}{l}
		\bar{\nu}_i \, \bar{t} \, t     \\
		\nu_i \, t \, \bar{t}     
	\end{array}\right. \, .  
\label{eq_3body_neutral}
\end{align}
Similar decays into a bottom- or light-flavor quark pair are suppressed by the small  
quark Yukawa couplings and contribute negligibly (given the assumption we have adopted of MFV).
\subsubsection{Two-body decays}
The decay width of a general two-body decay $0 \to 1~2$ follows directly 
from the spin and color averaged squared matrix element $|\overline{\mm}|^2$ ~\cite{kinematics},
\begin{align}
	\Gamma ( 0 \to 1~2) &= \frac{p_{cm}}{8\pi \,m_0^2} 
		\left| \overline{\mm} (0 \to 1~2) \right|^2 ,
\end{align}  
where $p_{cm}$ is the momentum of the final-state particles in the rest frame of 
the decaying particle,
$p_{cm}^2 = \frac{1}{4m_0^2} \left[ m_0^2 - (m_1 + m_2)^2 \right] \left[ m_0^2 - (m_1 - m_2)^2 \right]$.
  
Here we consider the decay of a fermionic octet $F_{\alpha}$ of mass $\mFal$ 
into a charged lepton (neutrino) and a charged (neutral) scalar octet state
of mass $\mSpm$ ($\mSRI$), \cf~\eqref{eq_2body}. We neglect the lepton mass in the decay rate calculations. 
Using the Feynman rules given in Appendix~\ref{sect_FeynmanRules}, one finds
\begin{align}
\begin{split}
	\left| \overline{\mm} (F_{\alpha} \to l_i^\mp\,S^{\pm}) \right|^2 &= 
		\frac{1}{8}\, \left| Y_{\nu}^{i\alpha} \right|^2 \, (\mFal^2 - \mSpm^2),
\\
	\left| \overline{\mm} (F_{\alpha} \to \nu_i\,S^0_{R/I}) \right|^2 &= 
		\frac{1}{16}\, \left| \left(V^T_{\rm PMNS} Y_{\nu}\right)_{i\alpha} \right|^2 \, (\mFal^2 - \mSRI^2),
\end{split}
\end{align}
and thus the partial decay rates are
\begin{align}
\begin{split}
	\Gamma ( F_{\alpha} \to l^\mp_i \,S^{\pm}) &= 
	\frac{1}{128 \pi } \,\left| Y_{\nu}^{i\alpha} \right|^2\, 
	\left( \mFal^2 - \mSpm^2 \right)^2\frac{1}{\mFal^3},
\\
	\Gamma ( F_{\alpha} \to \nu_i \,S^0_{R/I}) &= 
	\frac{1}{256 \pi } \,\left| \left(V^T_{\rm PMNS} Y_{\nu}\right)_{i\alpha} \right|^2\, 
	\left( \mFal^2 - \mSRI^2 \right)^2\frac{1}{\mFal^3}.
\end{split} 
\end{align}  
In the following we neglect the mass splitting between the scalar octet states, as favored by $b \to s \gamma$, and 
assume $m_S \equiv \mSpm \approx \mSR \approx \mSI$. 
The total two-body decay rate $\Gamma_{\rm 2, tot}$ of fermion $F_{\alpha}$ is
obtained by summing over all three lepton generations and by including the 
charge conjugated decays, 
\begin{align}
	\Gamma_{\rm 2, tot} (F_{\alpha})= \frac{1}{32\pi} \, 
	\sum_i \left| Y_{\nu}^{i\alpha} \right|^2 \, 
	(\mFal^2 - m_S^2)^2\, \frac{1}{\mFal^3}.
\end{align}
The neutrino mixing matrix $V_{\rm PMNS}$ cancels out in the sum over all three neutrino generations. 
The resulting two-body branching ratios are remarkingly simple 
and do not depend on the precise mass configuration of the fermion and scalar states,
\begin{align}
\begin{split}
	\BR (F_{\alpha} \to \ell_i^- S^+) + 
	\BR (F_{\alpha} \to \ell_i^+ S^-) &= \frac{1}{2} \, 
		\left| Y_{\nu}^{i\alpha} \right|^2 \, / \sum_i \left| Y_{\nu}^{i\alpha} \right|^2 , 
\\
 	\BR (F_{\alpha} \to \nu S^0) & = \frac{1}{2}, 
\label{eq_2bodyBRs_final}
\end{split}
\end{align}
where the sum of the branching ratios to charged leptons is then also one half.
\subsubsection{Three-body decays}
The decay width of a general three-body decay $0 \to 1~2~3$ 
can be parameterized as follows~\cite{kinematics},
\begin{align}
	\Gamma (0 \to 1~2~3) &= \frac{1}{(2\pi)^3} \frac{1}{32 m_0^3}
		\int_{p_{12}^{\rm min}}^{p_{12}^{\rm max}} \! dp_{12} \,
		\int_{p_{23}^{\rm min}}^{p_{23}^{\rm max}} \! dp_{23} \,
 		\left| \overline{\mm} (0 \to 1~2~3) \right|^2 ,
\end{align}  
with $p_{ij} = (p_i + p_j)^2$ and 
$p_i$ and $m_i$ denoting the momentum and the mass of particle $i$, $i= 0,1,2,3$.
The integration limits are given by,
\begin{align}
\begin{split}
	p_{12}^{\rm min} & = (m_1 + m_2)^2, \\
	p_{12}^{\rm max} & = (m_0 - m_3)^2, \\
	p_{23}^{\rm min/max} & = (E_2^* + E_3^*)^2 - 
		\left( \sqrt{E_2^{*^2} - m_2^2} \pm  \sqrt{E_3^{*^2} - m_3^2} \right)^2,
\end{split}
\end{align}
where 
$E_2^* =  (p_{12} - m_1^2 + m_2^2)/(2 \sqrt{p_{12}})$
and
$E_3^* = (m_0^2 - p_{12} - m_3^2)/(2 \sqrt{p_{12}})$ 
are the energies of particle 2 and 3 in the $p_{12}$ rest frame, respectively.

The squared matrix element for the three-body decay \eqref{eq_3body_charged} reads, 
in terms of the invariant $p_{23}$,     
\begin{align}
	\left| \overline{\mm} (F_{\alpha} \to \ell_i^- \,t \,\bar{b}) \right|^2  &=
	\frac{1}{8}  \left| Y_{\nu}^{i\alpha} \right|^2 \, \left| V_{tb}\right|^2 \, 
	\frac{\eta_U^2\, m_t^2}{v^2} \,
	\frac{(\mFal^2 - p_{23}) ( p_{23} - m_t^2)}{(p_{23} - \mSpm^2)^2}. 
\end{align}
The three-body decay \eqref{eq_3body_neutral} can be mediated by an off-shell $S_R^0$ or $S_I^0$ scalar.  
The summed matrix element is obtained as follows,
\begin{align}
	\left| \overline{\mm} (F_{\alpha} \to \nu_i \,t\, \bar{t}) \right|^2  &=
	\frac{1}{16} \left| \left(V^T_{PMNS} Y_{\nu}\right)_{i\alpha} \right|^2 \, 
	 \frac{\eta_U^2 \, m_t^2}{v^2} \,
	(\mFal^2 - p_{23}) \,
	\left\lbrace
	\frac{ ( p_{23} - 4m_t^2)}{(p_{23} - \mSR^2)^2}
	+ 
	\frac{p_{23} }{(p_{23} - \mSI^2)^2}
	\right\rbrace.
\end{align}
For equal masses, $m_S = \mSRI$, the matrix element becomes,
\begin{align}
	\left| \overline{\mm} (F_{\alpha} \to \nu_i \,t\, \bar{t}) \right|^2  &=
	\frac{1}{8} \left| \left(V^T_{PMNS} Y_{\nu}\right)_{i\alpha} \right|^2 \, 
	 \frac{\eta_U^2 \, m_t^2}{v^2} \,
	\left\lbrace
	\frac{(\mFal^2 - p_{23}) ( p_{23} - 2m_t^2)}{(p_{23} - m_S^2)^2}
	\right\rbrace.
\end{align}
Since the top-quark mass cannot be neglected, the matrix elements and the partial widths for the 
charged lepton channel and the neutrino channel
have sligthly different kinematics. In contrast to the case of two-body decays, the  
resulting branching ratios for the three-body decays thus do have a residual dependence on the masses of the new color octet states.
This mass dependence is small however and for the study of the branching ratios we will focus on the flavor patterns as 
described by the two-body decays. It is important to mention that when the mass of the fermionic octets are in the range, 
$m_t \ < \ m_{F_\alpha} \ < \ 2 m_t$, the $F_\alpha$ can decay only into charged leptons
and the three-body decays into neutrinos is kinematically forbidden.
Therefore, in general, the above given results define lower bounds on the rates for decays into charged leptons.

\subsubsection{Numerical Results}

%
\begin{figure}[tb]
\plot{.55}{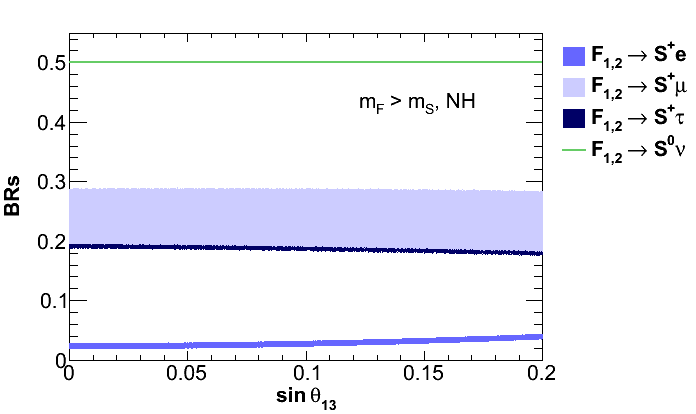}
\hspace*{-.11\linewidth}\plot{.55}{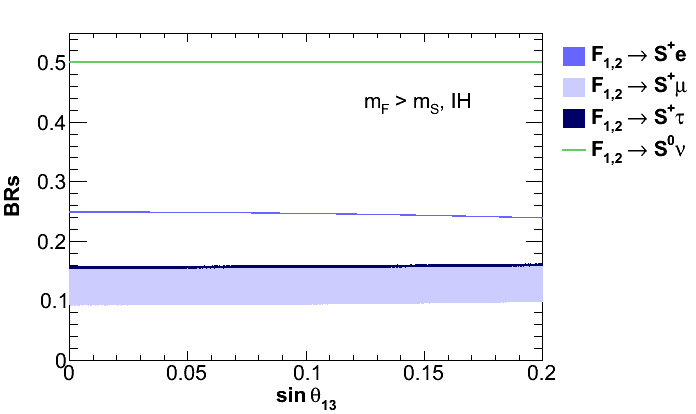}
\caption{Two-body branching ratios in the case of degenerate fermion masses
$m_{F1} = m_{F2}$ as a function of $\sin\theta_{13}$ both for normal hierarchy (NH, left) and inverted hierarchy (IH, right).
Full scan over the leptonic mixing parameters according to \eqref{eq_lept_mixingparams}. The bands for the  
$\mu$- and $\tau$-rates overlap, where the band width is mainly given by variations in $\sin\theta_{23}$.
\label{fig_BR2body_equalmass}
}
\end{figure}
We begin by focusing on the simplest case where the fermionic octets are degenerate in mass. 
In this scenario the branching ratios are independent of the unknown $\Omega$ matrix. Therefore, 
the results will only depend on the neutrino mixings and phases. In Fig.~\ref{fig_BR2body_equalmass} 
we show the total branching ratios versus $\sin\theta_{13}$ with a scan over all other neutrino parameters in the NH 
(left panel) and IH (right panel), respectively. Remarkably, the gross features of the branching ratios are independent 
of the neutrino parameters as is the value for the branching ratio to electrons.  This property can be understood
 by expressing the Yukawa couplings in terms of neutrino parameters (see Appendix~\ref{Yukawa.Neutrino}) and 
expanding in $s_{13}$.  For the NH:
\begin{align}
	\Gamma \left(F \to e X\right) & \propto \left(s_{12}^2 \sqrt{\Delta m^2_{21}} + s_{13}^2 \sqrt{\Delta m^2_{31}}\right),
	\\
	\Gamma \left(F \to \mu X\right) & \propto \left(c_{12}^2 \, c_{23}^2 \sqrt{\Delta m^2_{21}} + s_{23}^2 \sqrt{\Delta m^2_{31}}\right),
		\\
	\Gamma \left(F \to \tau X\right) & \propto \left(c_{12}^2 \, s_{23}^2 \sqrt{\Delta m^2_{21}} + c_{23}^2 \sqrt{\Delta m^2_{31}}\right).
\end{align}
From these expressions we see that in the electron branching ratios, the larger atmospheric mass is suppressed by $s_{13}^2$ whereas in the muon and tau channels this term has a factor of about a half (for tri-bimaximal mixing) leading to the order of magnitude hierarchy between these branching ratios.  This is a clear prediction of this degenerate fermionic mass scenario.  Furthermore, since the leading term in the electron branching ratio is the 
$s_{12}^2 \sqrt{\Delta m_{21}^2}$, which depends on more accurately measured parameters, the branching ratio itself can be predicted more accurately.

In the IH case one has:
\begin{align}
	\Gamma \left(F \to e X\right) & \propto \sqrt{\Delta m_{31}^2},
	\\
	\Gamma \left(F \to \mu X\right) & \propto c_{23}^2 \sqrt{\Delta m_{31}^2},
		\\
	\Gamma \left(F \to \tau X\right) & \propto s_{23}^2 \sqrt{\Delta m_{31}^2},
\end{align}
where the solar mass scale has been neglected. Here we see then that all branching ratios are proportional to the atmospheric mass scale with a slight suppression for the muon and tau channels, about a half for tri-bimaximal mixing.  Again the parameters in the electron branching ratio have small uncertainties leading to a more certain prediction for its value.  The upshot of this is that for the case of degenerate fermions, there is a definite prediction for the branching ratios, and measuring these would distinguish between the two possible neutrino spectra:
\begin{align}
	\BR \left(F \to \mu X \text{ or } \tau X\right) \gg \BR \left(F \to e X\right) & \Rightarrow \text{NH},
	\\
	\BR \left(F \to e X\right) >  \BR \left(F \to \mu X \text{ or } \tau X\right) & \Rightarrow \text{IH}.
\end{align}
\begin{figure}[tb]
\plot{.55}{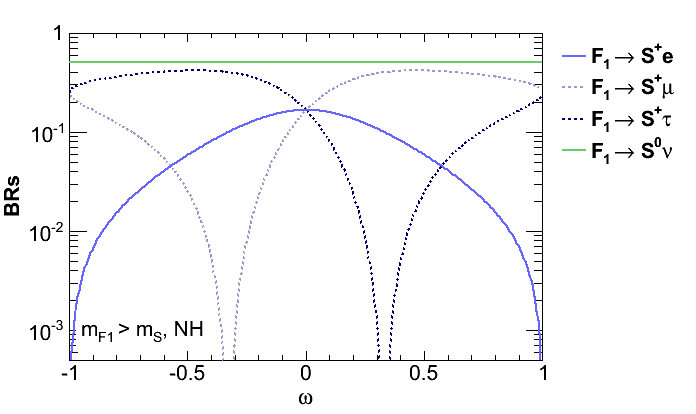}
\hspace*{-.12\linewidth}\plot{.55}{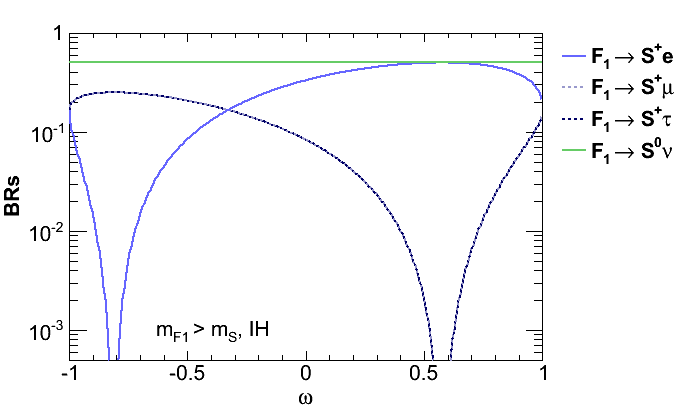}
\\[1ex]
\plot{.55}{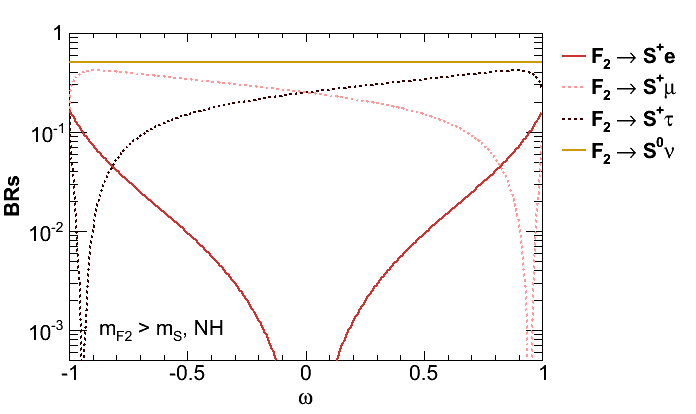}
\hspace*{-.12\linewidth}\plot{.55}{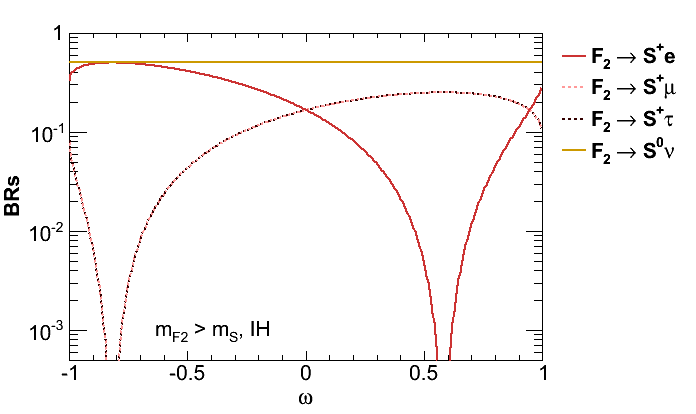}
\caption{Two-body branching ratios for $m_F > m_S$ as function of $\omega$ 
 both for normal hierarchy (NH, left) and inverted hierarchy (IH, right)
in the case of non-degenerate fermion masses.  Upper (lower) plots refer to $F_1$ ($F_2$) decays. 
Results do not depend on the precise mass configuration.
The parameter $\omega$ determines the neutrino Yukawa coupling, see \eqref{Omega} 
and \eqref{Dirac}. Central values for the neutrino mass differences and tribimaximal mixing in the leptonic 
mixing matrix are chosen without CP phases.    
\label{fig_BR2body_tribimax}
}
\end{figure}
For the more general scenario (non-degenerate $F$s) we start by assuming tri-bimaximal mixing in the neutrino sector for illustrative purposes, 
($\theta_{13}^{TB}=0$, $\sin\theta_{12}^{TB}=1/\sqrt{3}$, $\sin\theta_{23}^{TB}=1/\sqrt{2}$). 
The results are shown in Fig.~\ref{fig_BR2body_tribimax}, referring to $F_1$ ($F_2$) in the upper (lower) panels.
Already we see that the dependence on $\omega$ is quite important, even in this simplified case for neutrino mixing, 
since the branching ratios vary widely with $\omega$ and all of them go to zero at some specific $\omega$ value.  
However, general statements can still be made: 

\begin{itemize}

\item In the NH case either the $\mu$-channel or the $\tau$-channel always dominates over the $e$-channel.

\item In the IH the $\BR(F_{\alpha} \to \mu S)$ and $\BR(F_{\alpha} \to \tau S)$ coincide. 
The $e-$channel can be dominant for some values of $\omega$.

\end{itemize}

\begin{figure}[tb]
\plot{.55}{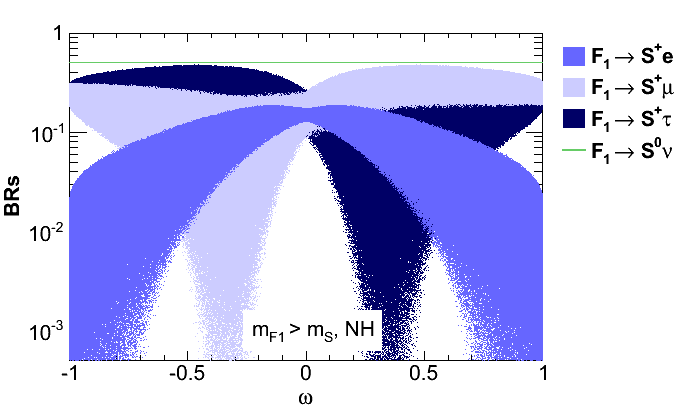}
\hspace*{-.12\linewidth}\plot{.55}{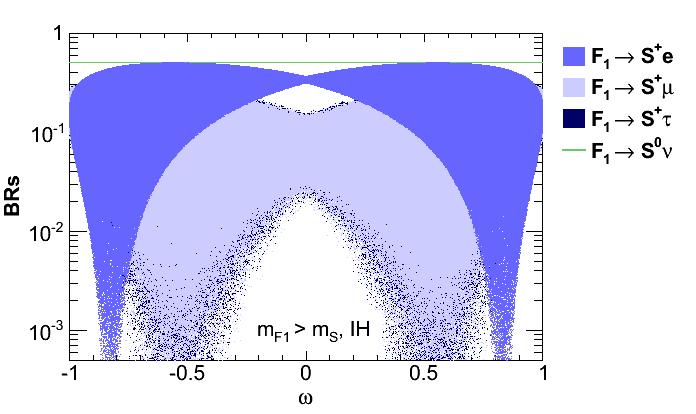}
\\[1ex]
\plot{.55}{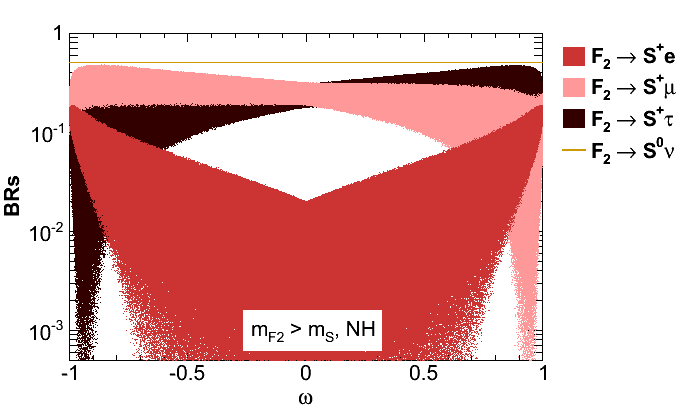}
\hspace*{-.12\linewidth}\plot{.55}{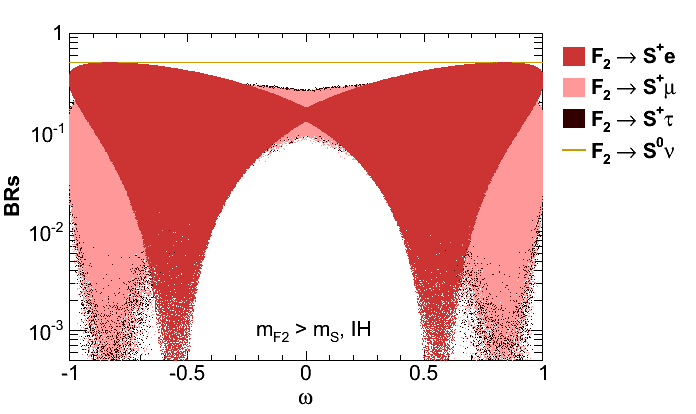}
\\[1ex]
\caption{Same plots as Fig.~\ref{fig_BR2body_tribimax}, but now the 
leptonic mixing parameters are varied between the experimental bounds given in \eqref{eq_lept_mixingparams}. 
Nonzero CP phases are included, $0 \leq \delta \leq 2\pi$. For the normal hierarchy, the majorana phase is varied
between  $0 \leq \Phi \leq \pi$ only for illustration purposes. 
\label{fig_BR2body_scan}
}
\end{figure}
Next we scan over all the neutrino parameters in order to understand their effect on the fermionic decays. We show the results in Fig.~\ref{fig_BR2body_scan} for the branching ratios of $F_1$ and $F_2$ versus $\omega$.  
This is of course even more complicated than Fig.~\ref{fig_BR2body_tribimax} and while we can still state that in the NH either the 
$\mu$-channel or the $\tau$-channel always dominates the $e$-channel,
we can no longer make the statement that in the IH the $\BR(F_{\alpha} \to \mu S)$ and $\BR(F_{\alpha} \to \tau S)$ coincide, due to varying $s_{23}$ and $s_{13}$ away from the tri-bimaximal values, which treat the muon and tau neutrinos equally.
It is important to mention that if this model is realized in nature and one finds the color octet fields at the LHC, 
and the neutrino spectrum is discovered in future neutrino experiments, then in principle one could 
determine the corresponding value of $\omega$ from the comparison with Fig.~\ref{fig_BR2body_scan}.

\subsection{Production Mechanisms and Lepton Number Violation}
In this section, we first present the cross section 
for the pair production of fermionic octets at the LHC. Second we will discuss the 
decay length. Finally, we combine our findings with the results of the decay properties. 
We focus on the signals for lepton number violation due to two final-state leptons of 
the same electric charge.

\subsubsection{Production of fermionic octets at the LHC}

\begin{figure}[tb] 
	\Fdiag{scale=.95}{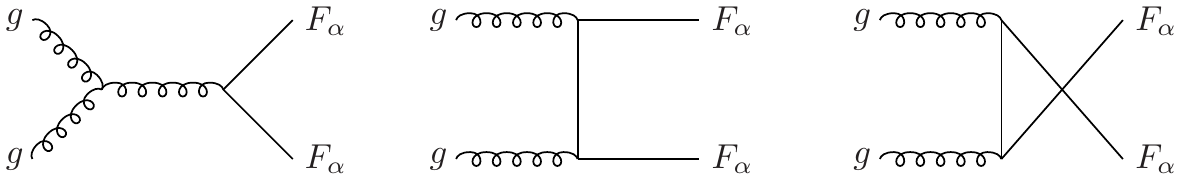} \hfill
	\Fdiag{scale=.95}{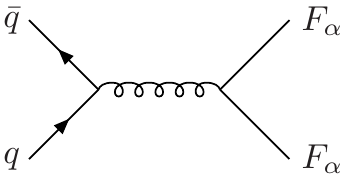}
\caption{
Parton-level Feynman diagrams for the pair production of color octet fermions at lowest order,
via gluon--gluon fusion (left) and quark--anti-quark annihilation (right).
\label{fig_feynman_prodFF}}
\end{figure}

The fermionic octets in the colored see-saw model can be produced at hadron colliders via the strong interaction. 
The dominant production mechanism proceeds at $\mathcal{O}(\alpha_s^2)$, 
via the two partonic channels gluon--gluon fusion and quark--anti-quark annihilation,
\begin{align}
g(p_1) ~~ g(p_2) ~\longrightarrow~ F_{\alpha} (p_3) ~~ F_{\alpha}(p_4) 
\qquad \text{and} \qquad
q(p_1) ~~ \bar{q}(p_2) ~\longrightarrow~ F_{\alpha} (p_3) ~~ F_{\alpha}(p_4),
\end{align}  
where $q=u,d,c,s,b$ can be any of the light-flavor quarks. 
The corresponding Feynman diagrams are shown in Fig.~\ref{fig_feynman_prodFF}. 
Mixed pairs of fermions, $F_1 F_2$, cannot be produced since the $gFF$ and 
$ggFF$ vertices are diagonal with respect to the fermionic eigenstates. We parameterize the cross sections
 in terms of the usual Mandelstam variables,
\begin{align}
\hat{s} = (p_1 + p_2)^2, \quad \hat{t} = (p_1 - p_3)^2, \quad \hat{u} = (p_1 - p_4)^2.
\end{align}
The differential partonic cross sections for the subprocesses 
are obtained from the spin- and color-averaged squared matrix elements,
\begin{align}
d\hat{\sigma}_{gg, \qqbar} (\hat{s}) = \frac{1}{16 \pi \hat{s}^2}
	 \left| \overline{\mm}_{gg, \qqbar}(\hat{s},\hat{t},\hat{u} )\right|^2 d\hat{t},
\end{align}   
which can be written as follows,
\begin{align}
\begin{split}
 \left| \overline{\mm}_{gg}(\hat{s},\hat{t},\hat{u}) \right|^2 &=
	18 \pi^2 \alpha_s^2 \, \left[1- \frac{1}{\hat{s}^2}(\hat{t}-\mFal^2)(\hat{u}-\mFal^2)\right]\,
\\& \times\,
	\left[ \frac{\hat{s}^2}{(\hat{t}-\mFal^2)(\hat{u}-\mFal^2)} - 2
	+ \frac{4 \mFal^2 \hat{s}}{ (\hat{t}-\mFal^2)(\hat{u}-\mFal^2)}\,
		\left( 1 - \frac{\mFal^2 \hat{s}}{ (\hat{t}-\mFal^2)(\hat{u}-\mFal^2)}\right) 
	\right],
\\[.5ex]
\left| \overline{\mm}_{\qqbar}(\hat{s},\hat{t},\hat{u}) \right|^2 &=
	\frac{64 \pi^2 \alpha_s^2}{3} \, \frac{1}{\hat{s}^2}
	\left[ 2 \mFal^2 \hat{s} + \left(\hat{t} - \mFal^2\right)^2 + \left(\hat{u} - \mFal^2\right)^2 \right].
\end{split}
\end{align}
Here, a factor $1/2$ has been taken into account because of the identical particles in the final
states. Note that the results for pair production of the fermionic states are the same as those for gluino pair production in supersymmetry with decoupled squarks~\cite{Beenakker:1996ch}.

At the hadronic level, the cross sections are obtained from the partonic ones by the convolution,
\begin{align}
	d\sigma_{PP \to F_{\alpha} F_{\alpha}}(s) = \int_{\tau_0}^1 \! d\tau \, 
	\left\lbrace
	\frac{d\mathcal{L}_{gg}^{PP}}{d\tau} \, d\hat{\sigma}_{gg}(\hat{s})
	+ \frac{d\mathcal{L}_{\qqbar}^{PP}}{d\tau} \, d\hat{\sigma}_{\qqbar}(\hat{s})
	\right\rbrace,
\end{align}
with $\tau = \hat{s}/s$, $s$ being the hadronic center-of-mass energy squared, and $\tau_0 = 4\mFal^2/s$ 
is the production threshold. The parton luminosities are given by,
\begin{align}
	\frac{d\mathcal{L}_{ab}^{AB}}{d\tau} = \frac{1}{1+\delta_{ab}} 
	\int_{\tau}^1\! \frac{dx}{x} \, \left[
		f_{a/A}(x,\mu) \, f_{b/B}\Big(\frac{\tau}{x},\mu\Big) 
	+ 	f_{b/A}\Big(\frac{\tau}{x},\mu\Big) \, f_{a/B}(x,\mu) \right],
\end{align}
where the parton distribution functions (PDFs) $f_{a/A}(x,\mu)$ parameterize the probability
of finding a parton $a$ inside a hadron $A$ with faction $x$ of the hadron momentum at 
a factorization scale $\mu$.

\begin{figure}[tb]
\plot{.55}{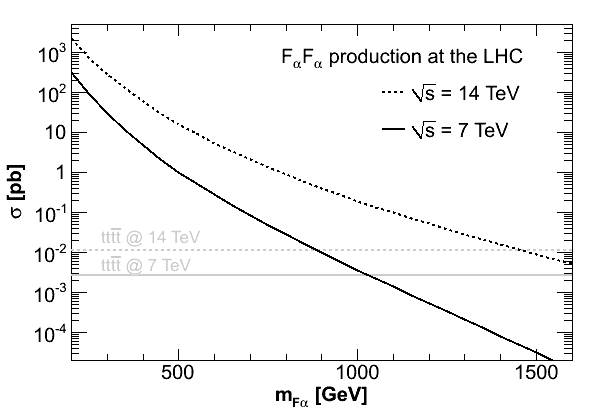}
\caption{
Hadronic cross section for the pair production of 
color octet fermions within the framework of the colored seesaw model
at the LHC with $\sqrt{s} = 7$\,TeV (solid) or $\sqrt{s} = 14$\,TeV (dashed), as a function of the mass of the produced fermion, $\mFal$. The cross sections for the dominant background, $tt\bar{t}\bar{t}$, are 
also shown, taken from~\cite{Barger:2010uw}.
\label{fig_crosssection}
}
\end{figure}

The numerical results are shown in Fig.~\ref{fig_crosssection}, both for $\sqrt{s} = 7$\,TeV 
and $\sqrt{s} = 14$\,TeV, as a function of the mass of the produced fermionic states. 
The PDF set CTEQ 6L~\cite{Pumplin:2002vw} is chosen 
and the factorization scale is set to $\mu = \mFal$.
Only for small momentum fractions $x$ and thus for light fermion masses, 
the $\qqbar$~channel is suppressed from the PDFs and 
less important than the $gg$~channel. At a 14\,TeV collider, the fraction of $\qqbar$ and $gg$ 
initial states is about 25\% and 75\%, respectively, for a fermion of mass $\mFal = 1$\,TeV. For 
$\sqrt{s} = 7$\,TeV the $\qqbar$~channel dominates over the $gg$~channel for $\mFal \gsim 1.2$\,TeV.    
The full hadronic cross section falls steeply with increasing fermion masses, from 
$\sigma \approx 300$\,pb for  $\mFal=200$\,GeV  to $\sigma \approx 1$\,fb for  $\mFal=1$\,TeV
in the early stage of the LHC at $\sqrt{s} = 7$\,TeV. For illustration we also show in Fig.~\ref{fig_crosssection} 
the cross section for the leading background,  $tt \bar{t} \bar{t}$, as we will discuss below. 

\subsubsection{Decay length and stable color octet fermions}

\begin{figure}[tb]
\plot{.49}{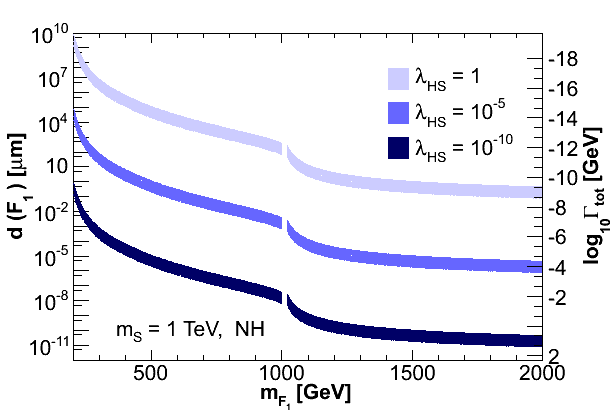}~
\plot{.49}{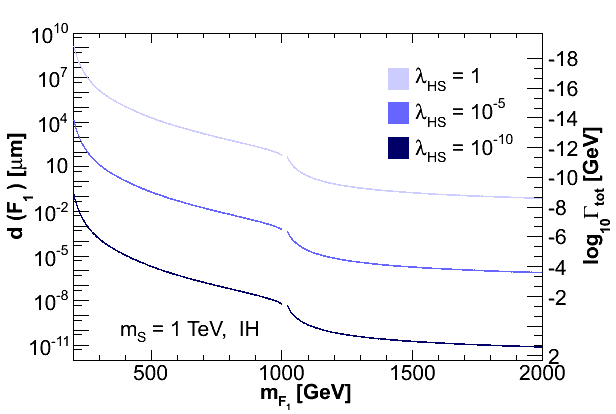}
\caption{
Decay length of the fermion for $m_S = 1$~TeV in case of normal hierarchy (NH, left) and inverted hierarchy (IH, right).
For $m_F < m_S$ the fermion decays via three-body decays, for $m_F > m_S$  the fermion undergoes two-body decays.
We scan over the parameter $-1 \leq \omega \leq 1$. The decays for $F_1$ and $F_2$ only differ for specific values of $\omega$ and agree when scanned over the full range of $\omega$.
\label{fig_length}
}
\end{figure}

The possible decays of the fermions depends on the ratio of the color octet masses,
 the neutrino Yukawa couplings, and the parameter $\lambda_{HS}$.
In Fig.~\ref{fig_length} we show their decay lengths, scanning over $-1 \leq \omega \leq 1$, for $m_S = 1$~TeV in case of normal hierarchy (NH, left in Fig.\ref{fig_length}) and inverted hierarchy (IH, right in Fig. \ref{fig_length}) versus the fermion mass.  Since we sum over all lepton generations, the dependence on the neutrino parameters disappears due to the unitarity of $V_{\text{PMNS}}$, hence there is no need to scan over these parameters.  This can be seen explicitly by expressing the Yukawas in terms of the neutrino parameters and $\omega$ and expanding in $s_{13}$ (see Appendix~\ref{Yukawa.Neutrino}).  The decay length in the NH and IH is then proportional to:
\begin{align}
\begin{split}
	d_{\text{NH}} & \propto \left[\sqrt{\Delta m_{21}^2} + \left(\sqrt{\Delta m_{31}^2}
	- \sqrt{\Delta m_{21}^2}\right) \omega^2\right]^{-1} \lambda_{HS},
\\
	d_{\text{IH}} & \propto \left[\sqrt{\Delta m_{31}^2}\right]^{-1}  \lambda_{HS},
\end{split}
\end{align}
respectively.  In the inverted hierarchy case it was possible to make a further expansion in the solar mass scale which led to the cancellation of the $\omega$ dependence at this order and is clearly reflected in Fig.~\ref{fig_length}.  Although the $\omega$ dependence is not strong in either spectrum.  For $m_F < m_S$ the fermion decays via three-body decays, for $m_F > m_S$  the fermion undergoes two-body decays.  It is clear from the plot that the parameter space allows for everything from prompt 
decays to displaced vertices and even decays outside the detector (large $\lambda_{HS}$ small fermion mass).

In the latter case where the fermions decay outside the detector, they exit the detector as so called R-hadrons, \textit{i.e.} they hadronize.  Such scenarios have been studied before, especially in  supersymmetry where the gluino can be (meta-)stable~\cite{Baer:1998pg}.  This scenario is analogous to ours  since the gluino's quantum number are the same as our $F$'s.  While such long lifetimes are discouraging for the observation of lepton number violation at colliders, the associated signals are still a spectacular indicator of new physics.

\subsubsection{Signals for lepton number violation}
The most important production mechanism for lepton number violation 
is the QCD pair production of fermionic octets. Each of these can subsequently decay into a charged lepton and either an on-shell or off-shell charged scalar octet which further decays into a top--bottom pair. In order to preserve the signature of same-sign dileptons, we must further require the top to decay purely hadronically into three jets, $t \to 3\,j$.  Then the signals of interest are~\cite{Perez-Wise}
\begin{align}
\begin{split}
	P \, P \ \to \ F_\alpha \, F_\alpha \  & \longrightarrow \ e^\pm \, e^\pm \ 8\,j;
\\
	& \longrightarrow \ \mu^\pm \ \mu^\pm \ 8\,j;
\\
	& \longrightarrow \ e^\pm \ \mu^\pm \ 8\,j.
\label{eq_decaychannels}
\end{split}
\end{align}
The tau channel is also very important to be help distinguish between the different neutrino spectra.
The event reconstruction will be much harder to carry out at the LHC. 
We continue by focusing on the simple case where the 
fermions are nearly degenerate therefore allowing more predictivity. The number of events $N$ per $10\,\text{fb}^{-1}$ 
for each of the signals is then given by
\begin{align}
N(\ell_i^{\pm} \ell_j^{\pm}) &=
	\Theta(m_{F_\alpha} - m_S) \,N_{\rm 2body}(\ell_i^{\pm} \ell_j^{\pm}) +
	\Theta(m_S - m_{F_\alpha}) \,N_{\rm 3body}(\ell_i^{\pm} \ell_j^{\pm}), 
\end{align}
with
\begin{align}
\begin{split}
N_{\rm 2body}(\ell_i^{\pm} \ell_j^{\pm}) =
	2 \sum_{\alpha = 1,2} &
	\sigma_{PP \to F_{\alpha} F_{\alpha}} \times
	\BR(F_{\alpha} \to \ell_i^\pm S^\mp) \times \BR(F_{\alpha} \to \ell_j^\pm S^\mp)
\\[-1.5ex]
	& \times \big[ \BR(S^\mp \to t b) \big]^2
	\times \big[\BR(t \to 3 \,j) \big]^2 \times 10
\\
	\approx 2 \sum_{\alpha = 1,2} & 
	\sigma_{PP \to F_{\alpha} F_{\alpha}} \times
	\BR(F_{\alpha} \to \ell_i^\pm S^\mp) \times \BR(F_{\alpha} \to \ell_j^\pm S^\mp)
	\times 4.6,
\end{split}
\end{align}
and
\begin{align}
\begin{split}
N_{\rm 3body}(\ell_i^{\pm} \ell_j^{\pm}) =
	2 \sum_{\alpha = 1,2} &
	\sigma_{PP \to F_{\alpha} F_{\alpha}}
	\times \BR(F_{\alpha} \to \ell_i^\pm t \, b) \times \BR(F_{\alpha} \to \ell_j^\pm t \, b)
	\times \big[\BR(t \to 3 \,j) \big]^2 \times 10 
\\
	\approx 2 \sum_{\alpha = 1,2} & \sigma_{PP \to F_{\alpha} F_{\alpha}}
	\times \BR(F_{\alpha} \to \ell_i^\pm t \, b) \times \BR(F_{\alpha} \to \ell_j^\pm t \, b)
	\times 4.6,
\end{split}
\end{align} 
where the first $\Theta$-function indicates two-body decays and the second three-body decays, the factor of $2$ comes from the sum over the charge conjugated decays, we have used $\BR(t \to 3 \,j) \sim 0.68$ and the cross section is in fb.  As an order of magnitude estimate, for $500$ GeV fermions and a branching ratio to muons at about $20 \%$ at the LHC for $7$ TeV center of mass energy one would expect about 180 like-sign dimuon events, before accounting for signal efficiency, which could play an important role.  
Requiring two isolated like-sign leptons and eight jets, the leading background comes from $t t \bar t \bar t$ production where one pair of like-sign tops decay semi-leptonically and the rest decay hadronically yielding the signal $l^\pm \, l^\pm 8\,j \, +$missing energy.  
The cross section for $t t \bar t \bar t$ is quite small, about $2$ ($11$)\,fb at $\sqrt{s} = 7$ ($14$)\,TeV~\cite{Barger:2010uw}.  
Addtitional background contributions arise from three tops plus jets, which is subleading  
due to the electroweak production mechanism ($0.5$\,fb at $14$\,TeV~\cite{Barger:2010uw}). 
$t \bar t b \bar b$ in comparison has a larger inclusive cross section (about $2600$\,fb at $14$\,TeV~\cite{Bredenstein:2009aj}), 
but the signal rate depends strongly on the fake rate for an isolated lepton resulting from a $b$ decay. 
Also, if we stick to the possibility that the top quarks can be reconstructed, see for example Ref.~\cite{Plehn}, 
and the simultaneous presence of two same-sign leptons,  $t \bar t b \bar b$ channel is not a relevant background 
in our study. Of course a more detailed study must be conducted to understand the observability. 

\begin{figure}[tb]
\plot{.49}{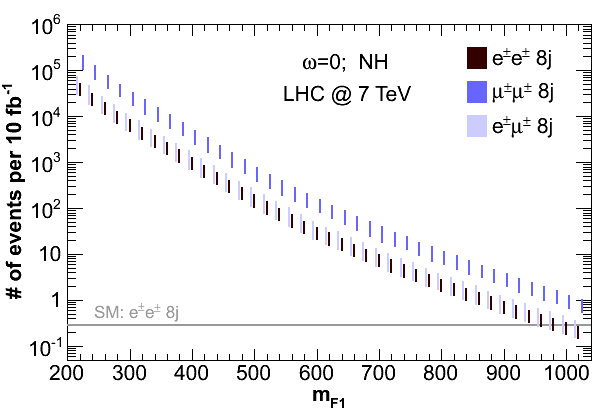}~
\plot{.49}{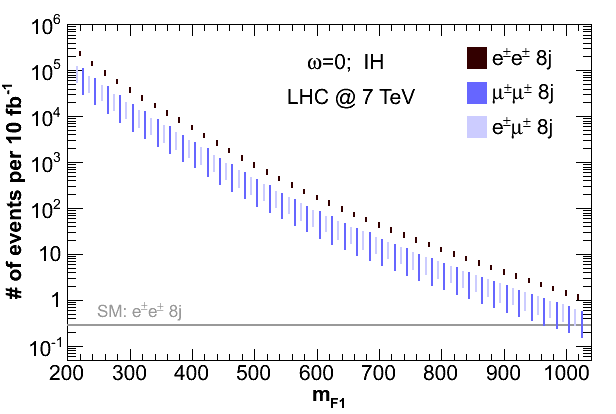}
\\[1ex]
\plot{.49}{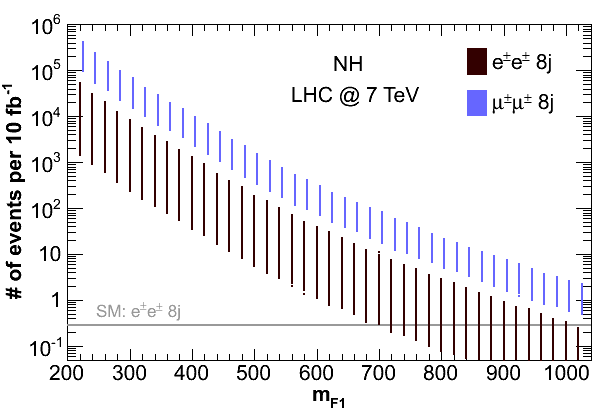}~
\plot{.49}{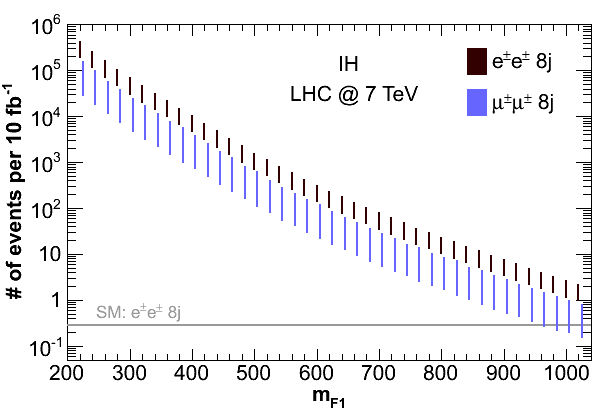}
\caption{
Rates for like-sign dilepton events per $10\,\text{fb}^{-1}$ from the pair production of color octet fermions
at the LHC with $\sqrt{s} = 7$\,TeV in the case of normal neutrino hierarchy (NH, left) and inverted hierachy (IH, right).
We scan over the leptonic mixing and $\omega=0$ (upper plots) or $-1\leq\omega\leq 1$ (lower plots). 
The fermions are assumed to be nearly degenerate in mass, $m_{F_1} \approx m_{F_2}$. The common scalar mass is set to $m_S = 2000$\,GeV for illustration purposes, but the results are not very sensitive on the scalar mass.
\label{fig_signal_sameMF}
}
\end{figure}

\begin{figure}[tb]
\plot{.49}{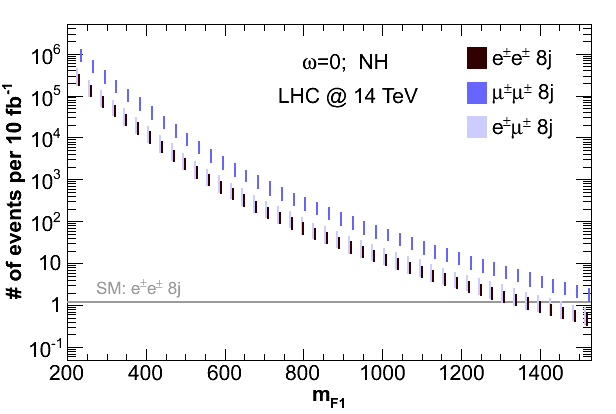}~
\plot{.49}{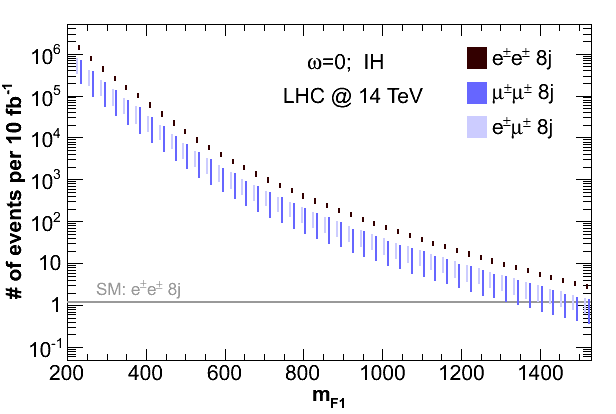}
\\[1ex]
\plot{.49}{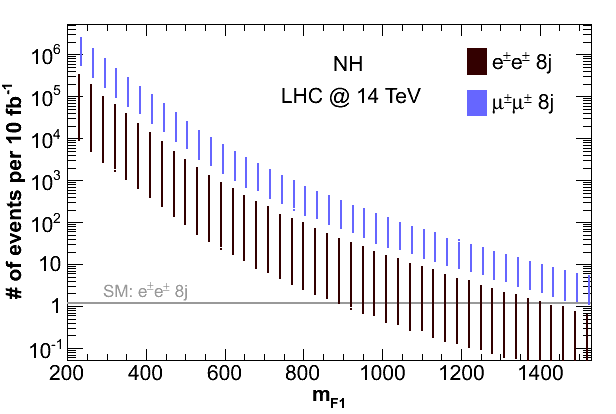}~
\plot{.49}{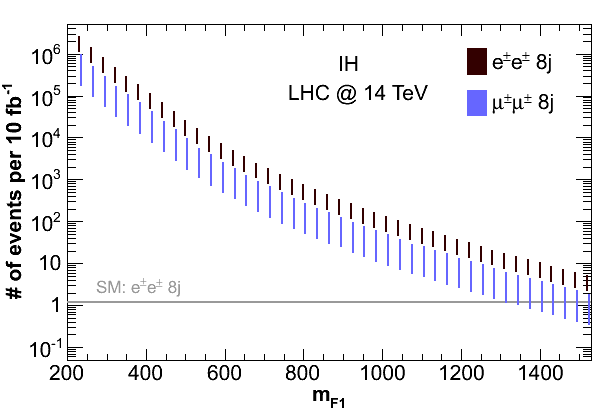}
\caption{
Same as Figure~\ref{fig_signal_sameMF} but for the LHC with $\sqrt{s} = 14$\,TeV.
\label{fig_signal_sameMF_14TeV}
}
\end{figure}


We present the signal and background in Figs.~\ref{fig_signal_sameMF} 
and \ref{fig_signal_sameMF_14TeV} at the LHC with $\sqrt{s} = 7$ and 14 TeV, respectively.
In both panels, the bands are due to the scan over the neutrino parameters.
The upper panels are with the choice of $\omega=0$, 
and in the lower panels we scan over $-1 \leq \omega \leq 1$ to indicate the possible range
of the signal rate. 
The common scalar mass is set to $m_S = 2000$\,GeV, \ie~the fermions decay via three-body decays. 
As explained above, the results for parameter regions with two-body decays look 
very similar. We see that the number of signal events is larger than the SM background up to fermion masses of about $1$ TeV for the dimuon channel at this level of analysis.

The rates can be understood by studying the results of Fig.~\ref{fig_BR2body_tribimax} (and~\ref{fig_BR2body_scan}).
For the specific case of $\omega =0$, for the normal neutrino hierarchy (left), the electron and the muon 
rate of $F_1$ decays are identical while the electron channels for $F_2$ decays are zero. As a consequence, 
only the pair production of $F_1 F_1$ can lead to $e^{\pm} e^{\pm}$ and $e^{\pm} \mu^{\pm}$  signals, and they have 
the same rate, while both   $F_1 F_1$  and  $F_2 F_2$ contribute to $\mu^{\pm} \mu^{\pm}$ signals.  
For the case of inverted hierarchy (right) and $\omega=0$, $F_2$~production contributes equally to $e$ and $\mu$ rates and $F_1$ predominantly to $e$ signals.

In both the top panels and the bottom panels, the neutrino spectrum has left its mark on the ratios of the $e\, e$ and $\mu \, \mu$ number of events.  Specifically in the normal hierarchy the number of muon events is larger than the number electrons events and vice versa in the inverted hierarchy.  Then we have verified the earlier statement that in addition to seeing the like-sign dileptons, in the special case where the fermions are degenerate comparing the number of muon events to the number of electron events can also reveal the neutrino spectrum.

So far, we have not included any experimental acceptance. The additional cuts  will undoubtedly reduce the signal rate
by a significant factor. On the other hand, the kinematical requirements in the events, 
such as no large missing energies, mass peaks
at $m_{F_{\alpha}} \approx m(\ell t b)$ and possibly at $m_{S_{\pm}} \approx m(t b)$, would help the signal identification and further background suppression as well. We leave the detailed simulations to future experimental
studies. For the study of the trilepton channels in see-saw models see Ref.~\cite{trileptons}.


\section{Summary}

There is an interesting possibility that 
the fields generating neutrino masses at one-loop level 
are scalar and Majorana fermionic color-octets of $SU(3)_{C}$, the colored seesaw mechanism. 
We reiterated the main features of this model, in comparison with the other common seesaw theories. 
We presented the bounds on certain model-parameters from the low energy decay $\mu \to e\gamma$
and from the observed neutrino mass spectrum and oscillation mixing parameters. 
We found that these parameters may have interesting correlations to the decay branching fractions 
of the colored states. 

In this context, we investigated the lepton number violating signals at the Large Hadron Collider
from the production and decay of the fermionic color-octet states.
Due to the QCD strong interaction, these states may be produced at the LHC with a favorable rate. 
We focused on the conclusive lepton number violating channel: same-sign dileptons (muons and electrons) channels.  
We found that for fermionic octets with mass up to about 1 TeV, the number of same-sign 
dilepton plus multiple jets events is larger than the SM background, indicating that this might 
be a very promising signal at the LHC which deserves a more detailed study.  
One may be able to use the fermionic octet decays of the lepton flavor combinations to distinguish between the neutrino 
spectra. This becomes especially predictive in the simplest case  
when the two fermionic color-octets are degenerate in mass. 
Our results demonstrate the potential of the LHC in searching for new physics beyond the SM in connection with neutrino mass generation. 

\appendix

\section{Feynman Rules}
\label{sect_FeynmanRules}
\begin{figure} 
\Fdiag{scale=.95}{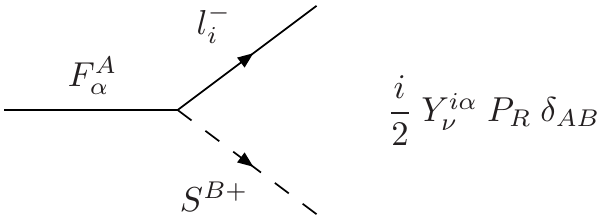} \hfill
\Fdiag{scale=.95}{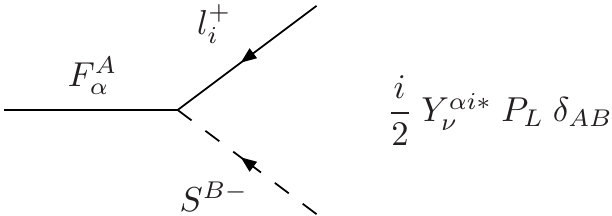} \hspace*{2cm}
\\[2ex]
\Fdiag{scale=.95}{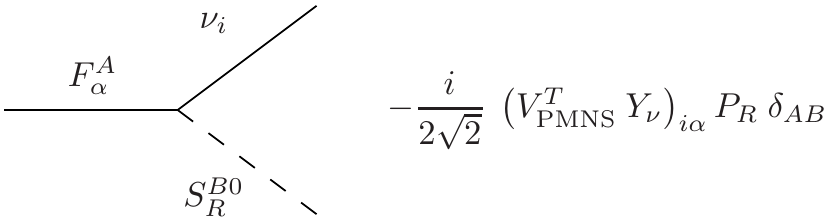} \hfill
\Fdiag{scale=.95}{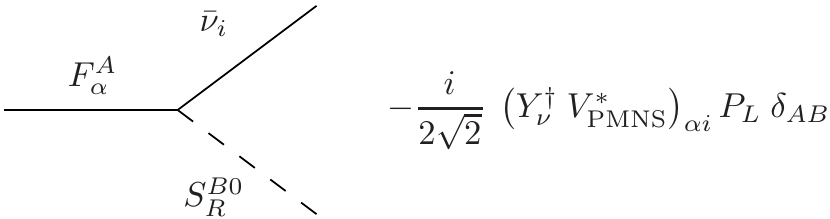}
\\[2ex]
\Fdiag{scale=.95}{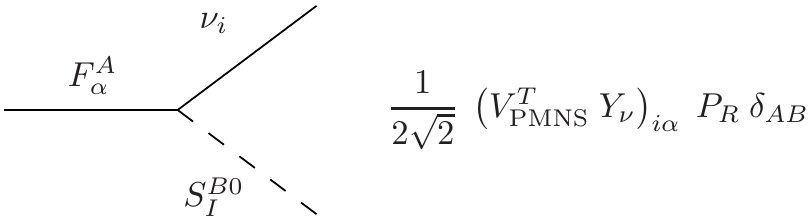} \hfill
\Fdiag{scale=.95}{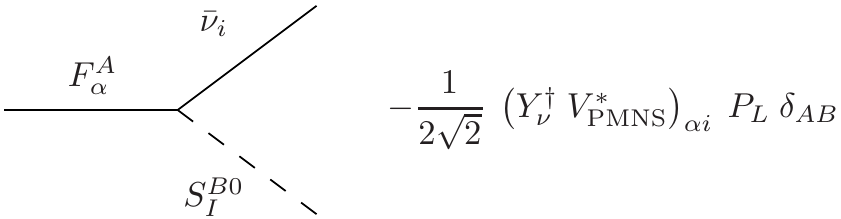}
\caption{Feynman rules for the Yukawa interactions of the fermionic octet.
The factor of one half results from the trace of the $SU(3)$ generators.
\label{fig_Feynman_F}}
\end{figure}

\begin{figure}
\Fdiag{scale=.95}{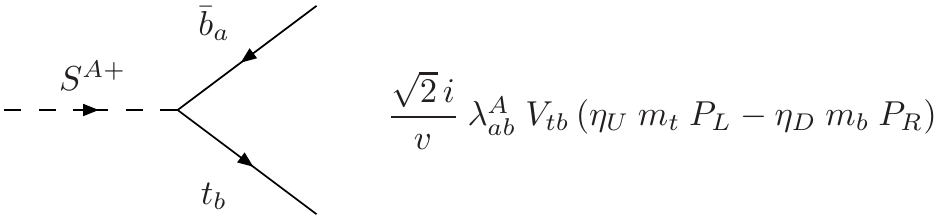}  
\\[2ex]
\Fdiag{scale=.95}{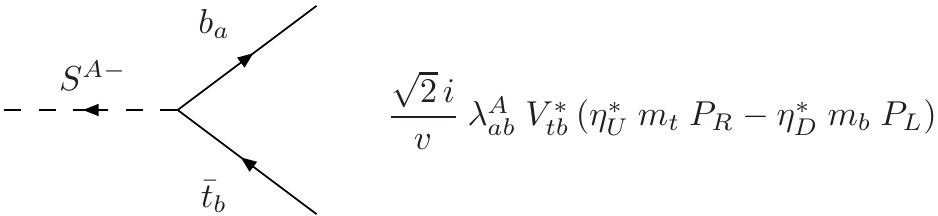} 
\\[2ex]
\hspace*{1cm}\Fdiag{scale=.95}{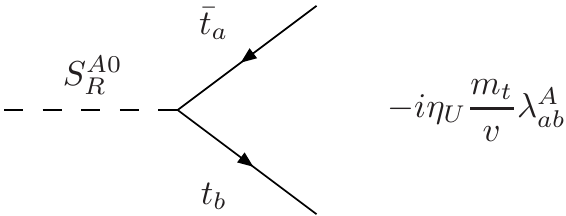}  \hspace*{2.2cm} 
\Fdiag{scale=.95}{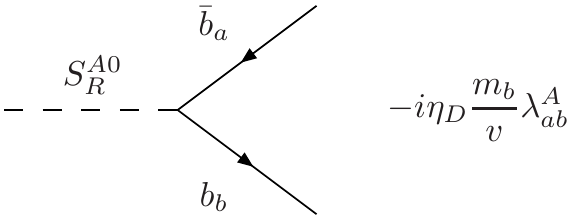}   \hspace*{1cm} 
\\[2ex]
\hspace*{1cm}\Fdiag{scale=.95}{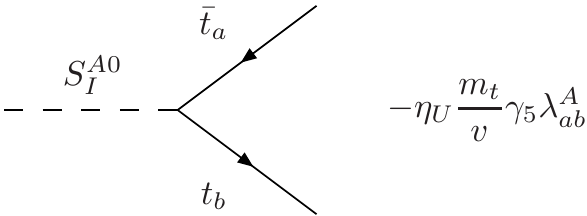}   \hspace*{2cm} 
\Fdiag{scale=.95}{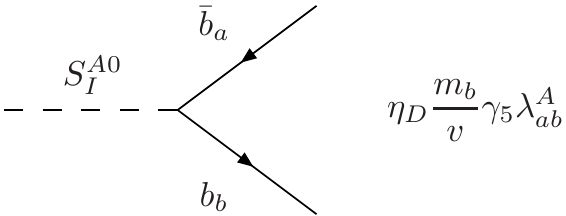}  \hspace*{1cm} 
\caption{Feynman rules for the scalar octet. 
\label{fig_Feynman_S}}
\end{figure}

%
The Feynman rules for the Yukawa interactions of the fermionic octet are collected  
in Figure~\ref{fig_Feynman_F} and those for the scalar octet in Figure~\ref{fig_Feynman_S}.
Here $\alpha = 1..2$ indicates the different generations of $F$, $i=1..3$ the generations of the leptons, and
upper (lower) case roman letters denotes indices of the adjoint (fundamental) representation of $SU(3)$.
$V_{tb}$ is the corresponding entry of the quark mixing matrix $V_{\rm CKM}$.


\section{Yukawa Couplings}
\label{Yukawa.Neutrino}
As indicated in Eq.~(\ref{Dirac}) the Yukawa couplings can be expressed in terms of the neutrino parameters and the matrix $\Omega$.  This can be illuminating for calculating various branching ratios and we do so explicitly here where we have expanded in $s_{13}$.  We furthermore use Eq.~\ref{eq_Meff} to express the results in terms of $\lambda_{HS}$.  In the following, $I_{\alpha} \equiv I(m_{\rho_{\alpha}}, m_S)$ where $\alpha = 1..2$ (the dimensionality of $I_{\alpha}$ is inverse mass).

\noindent \textbf{Normal Hierarchy}
\begin{align}
\begin{split}
	Y_\nu^{11} & = \frac{4 \pi}{v \, \sqrt{\lambda_{HS} \, I_1}}
				\left(
					\sqrt{1 - \omega^2} \,s_{12}  \, e^{i \frac{\Phi}{2}} \left(\Delta m_{21}^2\right)^{1/4}
					+ \omega \, s_{13} \, e^{-i \delta}
						\left(\Delta m_{31}^2\right)^{1/4}
				\right)
	\\
	Y_\nu^{12} & = -\frac{4 \pi}{v \, \sqrt{\lambda_{HS} \, I_2}}
				\left(
					\omega \, s_{12} \, e^{i \frac{\Phi}{2}} \left(\Delta m_{21}^2\right)^{1/4}
					- \sqrt{1 - \omega^2} \, s_{13} \,e^{-i \delta}
						\left(\Delta m_{31}^2\right)^{1/4}
				\right)
	\\
	Y_\nu^{21} & = \frac{4 \pi}{v \, \sqrt{\lambda_{HS} \, I_1}}
				\left(
					\sqrt{1 - \omega^2} \, c_{12} \, c_{23} \, e^{i \frac{\Phi}{2}}
						\left(\Delta m_{21}^2\right)^{1/4}
					+ \omega \, s_{23} \left(\Delta m_{31}^2\right)^{1/4}
				\right)
	\\
	Y_\nu^{22} & = -\frac{4 \pi}{v \, \sqrt{\lambda_{HS} \, I_2}}
				\left(
					\omega \, c_{12} \, c_{23} \, e^{i \frac{\Phi}{2}}
						\left(\Delta m_{21}^2\right)^{1/4}
					- \sqrt{1 - \omega^2} \, s_{23} \left(\Delta m_{31}^2\right)^{1/4}
				\right)
	\\
	Y_\nu^{31} & = -\frac{4 \pi}{v \, \sqrt{\lambda_{HS} \, I_1}}
				\left(
					\sqrt{1 - \omega^2} \, c_{12} \, s_{23} \, e^{i \frac{\Phi}{2}}
						\left(\Delta m_{21}^2\right)^{1/4}
					- \omega \, c_{23} \left(\Delta m_{31}^2\right)^{1/4} 
				\right)
	\\
	Y_\nu^{32} & = \frac{4 \pi}{v \, \sqrt{\lambda_{HS} \, I_2}}
				\left(
					\omega \, c_{12} \, s_{23} \, e^{i \frac{\Phi}{2}} \left(\Delta m_{21}^2\right)^{1/4}
					+ \sqrt{1 - \omega^2} \, c_{23} \left(\Delta m_{31}^2\right)^{1/4}
				\right)
\end{split}
\end{align}

The partial widths (and branching ratios) for the decays of the octet fermions are proportional to Yukawa couplings squared.  Since the Yukawa couplings are the sums of two terms with different dependence on $\omega$, they always cancel at some value of $\omega$ for a zero phase.  The total width involves a sum over all three lepton generations and as such is independent of the $V_{\text{PMNS}}$ parameters:
\begin{align}
\begin{split}
	\Gamma\left(F_1 \to X\right) & \propto
		\frac{16 \pi^2}{\lambda_{HS} \, I_1 \, v^2}
		\left(
			\sqrt{\Delta m_{21}^2}
			+ \left(
				\sqrt{\Delta m_{31}^2} - \sqrt{\Delta m_{21}^2}
			\right) \omega^2
		\right)
	\\
	\Gamma\left(F_2 \to X\right) & \propto
		\frac{16 \pi^2}{\lambda_{HS} \, I_2 \, v^2}
		\left(
			\sqrt{\Delta m_{21}^2}
			+ \left(
				\sqrt{\Delta m_{31}^2} - \sqrt{\Delta m_{21}^2}
			\right) \omega^2
		\right)
\end{split}
\end{align}
The above are valid for both two and three body decays.  Varying $\omega$ changes the total width by a little less than an order of magnitude, due to same factor between the atmospheric and solar mass scales.

In the case of degenerate octet fermions masses, $I_{1,2} \equiv I$,
 and $\Omega \in \mathcal{R}$ there is no dependence on $\omega$,
\begin{align}
\begin{split}
	\Gamma\left(F \to e^- X\right) & \propto
		\frac{16 \pi^2}{\lambda_{HS} \, I \, v^2}
		\left(
			s_{12}^2 \sqrt{\Delta m_{21}^2}
			+ s_{13}^2 \sqrt{\Delta m_{31}^2}
		\right)
	\\
	\Gamma\left(F \to \mu^- X\right) & \propto
	\frac{16 \pi^2}{\lambda_{HS} \, I \, v^2}
		\left(
			c_{12}^2 c_{23}^2 \sqrt{\Delta m_{21}^2}
			+ s_{23}^2 \sqrt{\Delta m_{31}^2}
		\right)
	\\
	\Gamma\left(F \to \tau^- X\right) & \propto
	\frac{16 \pi^2}{\lambda_{HS} \, I \, v^2}
		\left(
			c_{12}^2 s_{23}^2 \sqrt{\Delta m_{21}^2}
			+ c_{23}^2 \sqrt{\Delta m_{31}^2}
		\right)
\end{split}
\end{align}
Again, this is true for two or three body decays.  The two terms in the electron branching ratio are both suppressed, one by the solar mass scale and the other by the sin of $\theta_{13}$ so that the other two branching ratios dominate due to the atmospheric mass term.  These two branching ratios are equal in the tri-bimaximal case since $c_{23} = s_{23}$. 

Note that in the degenerated case, the branching ratio for the rare decay $\mu \to e \gamma$ is also independent of $\omega$,
\begin{equation}
	\BR\left(\mu \to e \gamma \right) = 
	\frac{192 \pi^3 \alpha_\text{EM} \mathcal{F}^2(x)}
	{\lambda_{HS}^2 I^2 G_F^2 v^4 m_S^4}
	\left|
		c_{12} \, c_{23} \, s_{12} \sqrt{\Delta m_{21}^2}
		+ s_{13} \, s_{23} \, e^{-i\left(\delta - \frac{\Phi}{2}\right)} \sqrt{\Delta m_{31}^2}
	\right|^2,
\end{equation}
which goes to zero at $s_{13} \sim 0.09$ and $\delta - \frac{\Phi}{2}=\pi$.

\noindent \textbf{Inverted Hierarchy}
\begin{align}
\begin{split}
	Y_\nu^{11} & = \frac{4 \pi}{v \, \sqrt{\lambda_{HS} \, I_1}}
				\left(
					\omega \, s_{12} \, e^{i \frac{\Phi}{2}}
						\left(\Delta m_{21}^2 + \Delta m_{31}^2\right)^{1/4}
					+ \sqrt{1 - \omega^2} \, c_{12} \left(\Delta m_{31}^2\right)^{1/4}
				\right)
	\\
	Y_\nu^{12} & = -\frac{4 \pi}{v \, \sqrt{\lambda_{HS} \, I_2}}
				\left(
					\sqrt{1 - \omega^2} \,  s_{12} \, e^{i \frac{\Phi}{2}}
						\left(\Delta m_{21}^2 + \Delta m_{31}^2\right)^{1/4}
					- \omega \, c_{12} \,  \left(\Delta m_{31}^2\right)^{1/4}
				\right)
	\\
	Y_\nu^{21} & = c_{23} \, \frac{4 \pi}{v \, \sqrt{\lambda_{HS} \, I_1}}
				\left(
					\omega \, c_{12} \, e^{i \frac{\Phi}{2}}
						\left(\Delta m_{21}^2 + \Delta m_{31}^2\right)^{1/4}
					- \sqrt{1 - \omega^2} \, s_{12} \left(\Delta m_{31}^2\right)^{1/4}
				\right)
	\\
	Y_\nu^{22} & = c_{23} \, \frac{4 \pi}{v \, \sqrt{\lambda_{HS} \, I_2}}
				\left(
					\sqrt{1-\omega^2} \, c_{12} \, e^{i \frac{\Phi}{2}}
						\left(\Delta m_{21}^2 + \Delta m_{31}^2\right)^{1/4}
					+ \omega \, s_{12} \left(\Delta m_{31}^2\right)^{1/4}
				\right)
	\\
	Y_\nu^{31} & = - s_{23} \, \frac{4 \pi}{v \, \sqrt{\lambda_{HS} \, I_1}}
				\left(
					\omega \, c_{12} \, e^{i \frac{\Phi}{2}}
						\left(\Delta m_{21}^2 + \Delta m_{31}^2\right)^{1/4}
					-  \sqrt{1-\omega^2} \, s_{12}  \,  \left(\Delta m_{31}^2\right)^{1/4}
				\right)
	\\
	Y_\nu^{32} & = - s_{23} \, \frac{4 \pi}{v \, \sqrt{\lambda_{HS} \, I_2}}
				\left(
					\sqrt{1-\omega^2} \, c_{12} \, e^{i \, \frac{\Phi}{2}}
						\left(\Delta m_{21}^2 + \Delta m_{31}^2\right)^{1/4}
					+ \omega \, s_{12} \left(\Delta m_{31}^2\right)^{1/4}
				\right)
\end{split}
\end{align}
The total width in the inverted hierarchy is:
\begin{align}
\begin{split}
	\Gamma\left(F_1 \to X\right) & \propto
		\frac{16 \pi^2}{\lambda_{HS} \, I_1 \, v^2}
		\left(
			\sqrt{\Delta m_{31}^2} \left(1 - \omega^2\right)
			+  \sqrt{\Delta m_{31}^2 + \Delta m_{21}^2} \; \omega^2
		\right)
		\approx
		 \frac{16 \pi^2}{\lambda_{HS} \, I_1 \, v^2}
		 \sqrt{\Delta m_{31}^2}
	\\
	\Gamma\left(F_2 \to X\right) & \propto
		\frac{16 \pi^2}{\lambda_{HS} \, I_2 \, v^2}
		\left(
			\sqrt{\Delta m_{31}^2} \; \omega^2
			+  \sqrt{\Delta m_{31}^2 + \Delta m_{21}^2} \left(1-\omega^2\right)
		\right)
		\approx
		 \frac{16 \pi^2}{\lambda_{HS} \, I_2 \, v^2}
		 \sqrt{\Delta m_{31}^2}
\end{split}
\end{align}
so that varying $\omega$ does not much influence on the width.

For degenerate octet fermions masses and $\Omega \in \mathcal{R}$:
\begin{align}
\begin{split}
	\Gamma\left(F \to e^- X\right) & \propto \frac{16 \pi^2}{\lambda_{HS} \, I \, v^2}
		\left(
			c_{12}^2 \sqrt{\Delta m_{31}^2}
			+ s_{12}^2 \sqrt{\Delta m_{31}^2 + \Delta m_{21}^2}
		\right) \approx \frac{M^\text{\text{Eff}}}{v^2} \sqrt{\Delta m_{31}^2}
	\\
	\Gamma\left(F \to \mu^- X\right) & \propto \frac{16 \pi^2}{\lambda_{HS} \, I \, v^2}
		\left(
			s_{12}^2 \, c_{23}^2 \sqrt{\Delta m_{31}^2}
			+ c_{12}^2 \, c_{23}^2 \sqrt{\Delta m_{31}^2 + \Delta m_{21}^2}
		\right) \approx \frac{M^\text{\text{Eff}}}{v^2} c_{23}^2 \sqrt{\Delta m_{31}^2}
	\\
	\Gamma\left(F \to \tau^- X\right) & \propto \frac{16 \pi^2}{\lambda_{HS} \, I \, v^2}
		\left(
			s_{12}^2 \, s_{23}^2 \sqrt{\Delta m_{31}^2}
			+ c_{12}^2 \, s_{23}^2 \sqrt{\Delta m_{31}^2 + \Delta m_{21}^2}
		\right) \approx \frac{M^\text{\text{Eff}}}{v^2} s_{23}^2 \sqrt{\Delta m_{31}^2}
\end{split}
\end{align}
Here there is a very mild hierarchy between the electron branching ratio and the muon and tau do to $s_{23} \sim c_{23} \sim 0.5$ and this is the only parameter that plays an important role.  Again the muon and tau branching ratios are equal in the tri-bimaximal case.

The branching ratio for $\mu \to e \gamma$ is a bit tricky in the inverted hierarchy since it vanishes in the zeroth order expansions used above.  Therefore, one must use the full PMNS matrix:
\begin{equation}
	\BR\left(\mu \to e \gamma \right) = 
	\frac{192 \pi^3 \alpha_\text{EM} \mathcal{F}^2(x)}
	{\lambda_{HS}^2 I^2 G_F^2 v^4 m_S^4}
	\left|
		-s_{13} \, s_{23} \, e^{-i \delta} \sqrt{\Delta m_{31}^2}
		+ \frac{1}{2} c_{12} \, s_{12} \, c_{23} \, \frac{\Delta m_{21}^2}{\sqrt{\Delta m_{31}^2}}
	\right|^2.
\end{equation}
This goes to zero at around $s_{13} \sim 0.007$ and $\delta=0$.

\subsection*{Acknowledgment}
P.F.P. would like to thank Mark B. Wise for discussions.
This work was supported in part by the US DOE under contract No.~DE-FG02-95ER40896.

\end{document}